\documentclass[aps,prd,reprint,groupedaddress,nofootinbib]{revtex4-2}

\makeatletter
\let\@fnsymbol\@arabic
\makeatother

\usepackage{graphicx,amssymb,amsmath,amsthm,amsfonts,epsfig,epsf,array,subfigure}
\usepackage[usenames]{color}
\usepackage{xcolor}
\usepackage{booktabs}
\usepackage{comment}
\usepackage{soul}
\usepackage{hyperref}

\def\be{\begin{equation}}
\def\ee{\end{equation}}
\def\bea{\begin{eqnarray}}
\def\eea{\end{eqnarray}}
\def\bfla{\begin{flalign}}
\def\efla{\end{flalign}}
\def\nn{\nonumber}

\def\gsim{\, \rlap{$>$}{\lower 1.1ex\hbox{$\sim$}}\,}
\def\lsim{\, \rlap{$<$}{\lower 1.1ex\hbox{$\sim$}}\,}

\def\ADM{M_{\text{tot}}}
\def\talpha{\widetilde{\alpha}}
\def\halpha{\widehat{\alpha}}

\definecolor{purple}{rgb}{0.7,0,1}

\begin{document}

\title{
Extreme mass ratio head-on collisions of black holes\\ in Einstein-scalar-Gauss-Bonnet theory}

\author{Antonia M. Frassino,$^{1,2,3}$ David~C.~Lopes,$^{4}$ and Jorge~V.~Rocha$^{5,6,7}$}

\affiliation{
$^1$SISSA, International School for Advanced Studies, via Bonomea 265, 34136 Trieste, Italy\\
$^2$INFN, Sezione di Trieste, via Valerio 2, 34127 Trieste, Italy\\
$^3$Departamento de F\'{i}sica y Matem\'{a}ticas, Universidad de Alcal\'{a}, Campus universitario 28805, Alcal\'a de Henares (Madrid), Spain\\
$^4$Centre for Astrophysics Research, University of Hertfordshire, College Lane, Hatfield AL10 9AB, UK\\
$^5$Departamento de Matem\'atica, ISCTE -- Instituto Universit\'ario de Lisboa, Avenida das For\c{c}as Armadas, 1649-026 Lisboa, Portugal\\
$^6$Departamento de F\'isica, Instituto Superior T\'ecnico -- IST, \\
Universidade de Lisboa - UL,
Avenida Rovisco Pais 1, 1049-001 Lisboa, Portugal\\
$^7$Instituto de Telecomunica\c{c}\~oes--IUL, Avenida das For\c{c}as Armadas, 1649-026 Lisboa, Portugal}

\date{\today}

\begin{abstract}
The evolution of the event horizon when two black holes merge can be determined by resorting to ray-tracing techniques on a single black hole spacetime, under the assumption that the binary's mass ratio is infinite and the underlying gravity theory respects the equivalence principle.
We extend this analysis to the head-on collision of non-spinning hairy black holes in Einstein-scalar-Gauss-Bonnet gravity.
In such theories the scalar field is coupled to a higher curvature operator, leading to possible modifications of the background geometry and consequently of photon propagation. 
We study three families of coupling functions: linear, quadratic, and a particular exponential form. The first choice enjoys a shift symmetry and forces the presence of scalar hair in the spectrum of black hole solutions. The latter two couplings break the shift symmetry and allow for spontaneously scalarized hairy black holes, which coexist with the Schwarzschild black hole. 
For all three classes of theories studied, we find a merger duration that is longer than the corresponding time in general relativity, when keeping the size of the small black hole fixed, and for viably small values of the coupling constant. However, the case of the exponential coupling yields a non-monotonic merger duration, which can become shorter than the general relativity value for a sufficiently large coupling constant.
We observe that the merger duration and the area increment generically track the behavior of the small black hole's photon ring.
Finally, we also compare our results with recent numerical simulations by other groups, despite the dissimilar mass ratios considered.
\end{abstract}



\maketitle

\section{Introduction\label{sec:Intro}}

The advent of gravitational wave (GW) astronomy has promoted coalescing compact binaries to precision probes of gravity in the strong-field, highly dynamical regime, making them natural laboratories to search for corrections to general relativity (GR) \cite{LISA:2022kgy,Barack:2018yly,Gnocchi:2019jzp,Baker:2014zba}. 
The LIGO–Virgo–KAGRA (LVK) network has already produced a rich catalogue of detections and stringent tests of GR \cite{LIGOScientific:2025snk,LIGOScientific:2025bkz}, and its sensitivity will improve in the coming years~\cite{2017arXiv170200786A, Punturo_2010, ET:2019dnz, Reitze:2019iox} delivering orders-of-magnitude gains in reach and measurement accuracy. 
Existing GW observations significantly constrain broad classes of deviations 
(see, for example, \cite{LIGOScientific:2025obp,LIGOScientific:2025rid}), 
indicating that very precise modeling and data analysis will be required to uncover potential departures from GR.

Among the many proposed extensions of GR, Einstein–scalar–Gauss–Bonnet (EsGB) gravity provides a particularly well-motivated and theoretically controlled framework.
In this class of theories, which we will focus on, a scalar field couples non-minimally to the Gauss–Bonnet invariant, introducing higher-curvature interactions while preserving second-order equations of motion in four dimensions. These theories support black hole (BH) solutions endowed with non-trivial scalar hair \cite{Kanti:1995vq, Sotiriou:2013qea, Sotiriou:2014pfa, Doneva:2017bvd, Silva:2017uqg, Antoniou:2017acq, Antoniou_2018_1}. The presence of the scalar field modifies both the dynamics and the radiative signature of compact binaries~\cite{Yagi:2011xp}, potentially altering their inspiral, merger, and ringdown. Yet, the theory can remain compatible with weak-field tests \cite{Amendola:2007ni, Elder:2022rak}. This makes strong-field GW observations of compact binaries a particularly powerful probe, provided their dynamics are modeled with sufficient accuracy.

Substantial effort has therefore been devoted to modeling GW signals in EsGB gravity, starting with early analytical studies~\cite{Yagi:2011xp} and numerical investigations under some approximations~\cite{Witek:2018dmd, Okounkova:2020rqw}.
In the context of EsGB gravity, confrontation of theoretical waveform models with observational data has led to increasingly stringent bounds on the Gauss–Bonnet coupling (see, e.g., \cite{Carson:2019fxr, Lyu:2022gdr,Sanger:2024axs,Wong:2022wni,Yordanov:2024lfk}), and EsGB-specific waveform models have been developed within the effective-one-body framework and applied to selected LVK events~\cite{Julie:2024fwy, Sanger:2024axs}. Beyond semi-analytical modeling, recent theoretical progress has established that higher-derivative scalar–tensor theories, including EsGB, admit a well-posed initial value formulation in the weak-coupling regime~\cite{Kovacs:2020pns, Figueras:2024bba}, enabling fully non-linear numerical simulations of compact binaries~\cite{East:2020hgw, East:2021bqk, Corman:2022xqg, Doneva:2023oww, AresteSalo:2025sxc, Corman:2025wun, Capuano:2026lhs} and direct comparisons with GW observations. Nevertheless, such simulations are computationally demanding and theory-specific, underscoring the importance of complementary approaches that enable scanning quickly a broad range of parameters. 
These considerations lead us to investigate mergers of scalar-hairy BHs in EsGB gravity from a different perspective.

In this paper, we perform a complementary study of BH mergers in EsGB gravity, by extending the analysis to the extreme mass-ratio (EMR) limit, where semi-analytical methods can be used. Specifically, we investigate head-on collisions of such hairy BHs when one companion is infinitely more massive than the other by resorting to the ray tracing technique developed in this context by Emparan and Mart\'inez~\cite{Emparan:2016ylg}. This approach is valid for theories respecting the weak equivalence principle, which is the case for EsGB gravity if matter is minimally coupled to the spacetime metric, and thus not coupled directly to the scalar field, nor to the Gauss-Bonnet invariant. Hence, in these theories photons follow geodesics, which is the key ingredient required to trace back in time the generators of the event horizon describing the merger. This technique has been employed in a number of more recent works~\cite{Emparan:2016ipc, Emparan:2017vyp, Emparan:2020uvt, Pina:2022dye, Dias:2023pdx, Gadioux:2024tlm, Dias:2024wib, Gadioux:2024jlx}.

We study three different families of EsGB theories for which quantitative results are obtained for the merger duration and area increase as a function of the scalar-Gauss-Bonnet coupling constant. 
We find that, when the scalar field is either linearly or quadratically coupled to the Gauss-Bonnet invariant, the highly non-linear merger phase proceeds slower than in GR, keeping the size of the small BH fixed, for comparison. The third class of models considered involves a particular exponential type of coupling that induces spontaneous scalarization of sufficiently small BHs (compared to the coupling constant). For this theory two distinct outcomes are possible: for sufficiently small values of the (dimensionless) coupling constant the merger is delayed when compared to GR, but there also exists a parameter range in which the merging time is accelerated, while not being ruled out by known instabilities of the BH geometry.

In the case of a linear coupling, our results align with those obtained for comparable mass binaries using numerical relativity~\cite{Corman:2025wun}, suggesting the phenomenon is a robust feature of shift-symmetric EsGB dynamics across a wide range of mass ratios.
This effect opposes the post-Newtonian expectation that the inspiral phase should be accelerated due to the additional emission of dipole scalar radiation and indicates that, in the strong field regime, the merger is delayed. 
In the case of the special exponential coupling, recent numerical simulations~\cite{Capuano:2026lhs} point in the opposite direction, finding shorter merging times for EsGB than in GR, which is also a possibility revealed by our analysis.

The structure of the paper is as follows. 
In Section~\ref{sec:EsGB} we review the literature concerning BH solutions in EsGB gravity. 
These geometries will be used in Section~\ref{sec:mergers} to construct the time evolution of the event horizon in EMR mergers. In this section we also include a brief description of how the ray tracing technique is implemented, and extract our main observables: the merger duration and the relative area increase of the small BH. We study three different families of EsGB theories for which we obtain quantitative results as a function of the scalar-Gauss-Bonnet coupling constant. 
Section~\ref{sec:comparison} is dedicated to the comparison with recent numerical relativity (NR) results for the case of the shift-symmetric theory and EsGB with a particular exponential form of the scalar-Gauss-Bonnet coupling function.
Finally, Section~\ref{sec:conclusion} is devoted to conclusions and some discussion. 
We relegate to three Appendices the details about the procedure to obtain the numerical solutions for the hairy black holes, and about the conversion of our results to units better adapted to NR simulations.

Throughout the manuscript we adopt geometrical units, in which Newton's constant of gravitation and the speed of light are set to unity, $G=c=1$.

\section{Black holes in Einstein-Scalar Gauss-Bonnet gravity\label{sec:EsGB}}

EsGB theories can be viewed as an extension of the effective field theory (EFT) approach to gravity~\cite{Donoghue:1994dn, Burgess:2003jk} to the case of scalar-tensor theories, in which the scalar-Gauss-Bonnet coupling appears naturally as the leading-order, higher-derivative correction to GR~\cite{Weinberg:2008hq}.
Such couplings are also characteristic of low-energy effective actions from string theory~\cite{Gross:1986mw, Metsaev:1987zx}.
In this language, the scalar–Gauss–Bonnet coupling plays the role of an EFT coefficient, and the resulting equations of motion are understood as a controlled low-energy expansion~\cite{Kovacs:2020pns,Figueras:2024bba}.

The action describing EsGB theory can be written as~\cite{Nojiri:2005vv, Antoniou:2017acq}
\be
\label{eq:Action}
{S} = \frac{1}{16 \pi} \int d^4 x \sqrt{-g} \left[ R - \frac{1}{2} \partial_a \Phi  \partial^a \Phi + \alpha f(\Phi) R_{\mathrm{GB}}^2 \right] \,, 
\ee
which contains the usual Einstein-Hilbert term, given by the Ricci scalar, $R$, as well as the Gauss-Bonnet term, defined by 
\be
R_{\mathrm{GB}}^2 = R^2- 4R_{a b}R^{a b}+R_{a b c d}R^{a b c d}\,,
\ee 
where $R_{a b}$ and $R_{a b c d}$ are the Ricci tensor and the Riemann tensor, respectively. Furthermore, the theory involves a scalar field, $\Phi$, which is coupled to the Gauss-Bonnet term through a generic function, $f(\Phi)$, which is rendered dimensionless upon pulling out the multiplicative coupling constant, $\alpha$, with dimensions of length squared.

Variation of the action with respect to the metric field, $g_{ab}$, and the scalar field, $\Phi$, yields the equations of motion for the theory, namely
\begin{eqnarray}
\label{eq:field_eqs_Phi}
\nabla_a \nabla^a \Phi + \alpha\dot{f}(\Phi) R_{\mathrm{GB}}^2 = 0 \,,\\
\label{eq:field_eqs_EGB}
R_{ab}-\frac{1}{2}R\, g_{ab} = T_{ab}^{\Phi} + \alpha\, T_{ab}^{\mathrm{GB}}\,, 
\end{eqnarray}
where $\nabla_a$ denotes the covariant derivative and the dot denotes a derivative with respect to $\Phi$. 
The stress-energy tensor has contributions both from the metric and the scalar field~\cite{Antoniou:2017acq},
\bea
    T_{ab}^{\Phi} \!\!&=&\!\! \frac{1}{2}\partial_a\Phi\partial_b\Phi - \frac{1}{4}g_{ab}\partial_c\Phi\partial^c\Phi\,,
    \\
    T_{ab}^{\mathrm{GB}} \!\!&=&\!\! \frac{1}{2g}\left( g_{ca}g_{db}-g_{da}g_{cb}\right) \epsilon^{edgh} \epsilon^{ckij} R_{ijgh} \nabla_k \nabla_e f,
\eea
where $\epsilon$ is the completely antisymmetric tensor in four dimensions.

Assuming a static and spherically symmetric spacetime with line element
\be
ds^2 = - e^{\Gamma (r)} dt^2 + e^{\Lambda (r)} dr^2 + r^2 (d\theta^2 + \sin^{2}{\theta}  \hspace{1mm} d\varphi^2)\,,
\ee
the metric field equations read 
\bea\label{eq:MFE1}
&&\!\!\!\!\!\!8 \alpha \dot{f} \left[\left(e^{\Lambda}\!-3\right) \Lambda' \Phi'-2 \left(e^{\Lambda}\!-1\right) \Phi ''\right] \! + 4 e^{\Lambda}\! \left(r \Lambda '+e^{\Lambda}\!-1\right)
\nn\\&&
=
\Phi '^2 \left[16 \left(e^{\Lambda }-1\right) \ddot{f}+r^2 e^{\Lambda }\right]\,,\\\nn \\
&&\!\!\!\!\!\!\label{eq:MFE2} r^2 e^{\Lambda } \Phi '^2+4 e^{\Lambda } \left(-r \Gamma '+e^{\Lambda }-1\right) =8 \left(e^{\Lambda }-3\right) \alpha\dot{f} \Gamma ' \Phi '\,, \\ \nn \\
&&\!\!\!\!\!\!\label{eq:MFE3} 
4 \alpha\dot{f} \left[-2 \Gamma ' \Phi ''-\Phi ' \left(2 \Gamma ''-3 \Gamma ' \Lambda '+\Gamma '^2\right)\right]
+e^{\Lambda } \big[2 r \Gamma '' \nn\\ &&  +\left(r \Gamma '+2\right) \left(\Gamma '-\Lambda '\right)\big]
=\!\Phi '^2 \left(8 \alpha \ddot{f} \Gamma ' -r e^{\Lambda }\right)\,,
\eea
and the scalar field equation reads
\bea\label{eq:SFE}
&&\!\!\!\!\!\!\!\!\!\!\!\!4 e^{-\Lambda } \alpha\dot{f} \left[\left(e^{\Lambda }-3\right) \Gamma ' \Lambda '-\left(e^{\Lambda }-1\right) \left(2 \Gamma ''+\Gamma '^2\right)\right]\nn\\
&&=-2 r^2 \Phi ''-r\Phi ' \left(r \Gamma '-r \Lambda '+4\right)\,,
\eea
where the primes denote derivatives with respect to $r$, and we omitted the functional dependence on $r$ and $\Phi$, i.e. $\Lambda=\Lambda(r)$, $\Gamma=\Gamma(r)$, $\Phi=\Phi(r)$ and $f=f(\Phi)$.

Depending on the choice of coupling function $f$ in Eq.~\eqref{eq:Action}, the EsGB theory may admit, or not, the same BH solutions as those of GR. When it does, the hairy BHs branch out from the GR solutions at specific points in the parameter space determined by the perturbative instabilities of the unscalarized BHs~\cite{Doneva:2017bvd}.
There is a countable infinity of such unstable points and so there are infinite branches of hairy black holes in such theories, labeled by $n\in\mathbb{N}_0$, the number of zeroes in the scalar field profile. These are so-called spontaneously scalarized BHs.
However, for a shift-symmetric theory, the GR BHs do not solve the equations of motion and the hairy BHs only exist along a single branch of solutions.

In this work, we focus on three different subclasses of EsGB gravity, distinguished by the choice of coupling functions. Specifically, we study coupling functions that are either linear in the scalar field~\cite{Sotiriou:2013qea}, quadratic~\cite{Silva:2017uqg}, or of a special exponential form~\cite{Doneva:2017bvd}. 
The latter two coupling functions satisfy $\dot{f}(0) = 0$, which ensures that the Schwarzschild spacetime, having $\Phi=0$, is also a solution of EsGB, along with additional hairy BHs.

In the remainder of this Section, we describe the BH solutions that arise in each of these theories. 
This is done by solving the system of equations~(\ref{eq:MFE1}-\ref{eq:SFE}), demanding regularity at the horizon, $r=r_h$, and imposing that the scalar field asymptotes to zero at infinity. The solutions obtained numerically fall along lines in the two-dimensional space parameterized by $\Phi(r_h) = \Phi_h$ and
\be
\talpha \equiv \frac{\alpha}{r_h^2}\,.
\label{eq:alphatilde_def}
\ee
More details about the numerical construction of the solutions are given in Appendix~\ref{app:f_solutions}.

\subsection{Linear coupling}

For our first choice of the coupling function we adopt a linear form, 
\be\label{eq:coupling_linear}
f(\Phi) = \frac{\Phi}{4}\,,
\ee
which gives rise to the shift-symmetric EsGB solutions obtained in~\cite{Sotiriou:2014pfa, Delgado:2020rev}. The theory in this case is invariant under $\Phi \to \Phi + c$, where $c$ is an arbitrary constant, and the normalization in Eq.~\eqref{eq:coupling_linear} is conventional. One can restrict to coupling constant $\alpha\geq0$ without loss of generality, since the equations of motion are invariant also under the $\mathbb{Z}_2$ transformation $(\alpha,\Phi)\to(-\alpha,-\Phi)$.

Shift-symmetric EsGB has attracted considerable interest. An important property of this theory is that the linear coupling forces static BHs to be hairy, and hence distinct from those of GR\footnote{This can be readily seen from Eq.~\eqref{eq:field_eqs_Phi}, as the coupling to the Gauss-Bonnet invariant never vanishes.}. In addition, previous studies have found that EsGB BH solutions with linear coupling are free from ghosts and Laplacian instabilities for small coupling~\cite{Minamitsuji:2022mlv}, and do not suffer from any quasinormal mode instabilities~\cite{Chung:2024vaf,Khoo:2024agm}, further motivating their study.

\begin{figure}[t!]
\includegraphics[width=7.5cm]{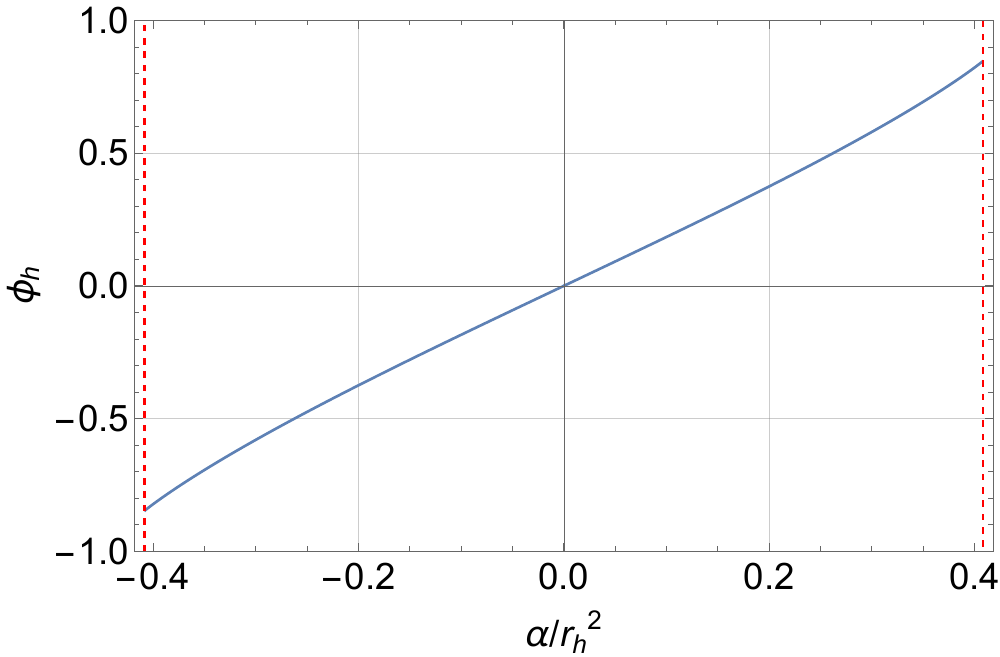}
\caption{Single branch of BH solutions of EsGB with linear coupling, $f(\Phi)=\Phi/4$, plotted in the ($\Phi_h$, $\alpha/r_h^2$) parameter space. These develop in the ($\Phi_h$, $\alpha/r_h^2$)-plane from the point ($\Phi_h=0$, $\alpha/r_h^2=0$) until Eq.~\eqref{eq:inequality} is no longer satisfied, i.e. for $|\alpha|/r_h^2 \geq 1/\sqrt{6}$, represented by the red-dashed curves. The Schwarzschild spacetime, with $\Phi=0$, is a solution of the theory only for $\alpha/r_h^2=0$.
\label{fig:solutions_linear_phi0_vs_alpha}}
\end{figure}

\begin{figure}[t!]
\includegraphics[width=7.5cm]{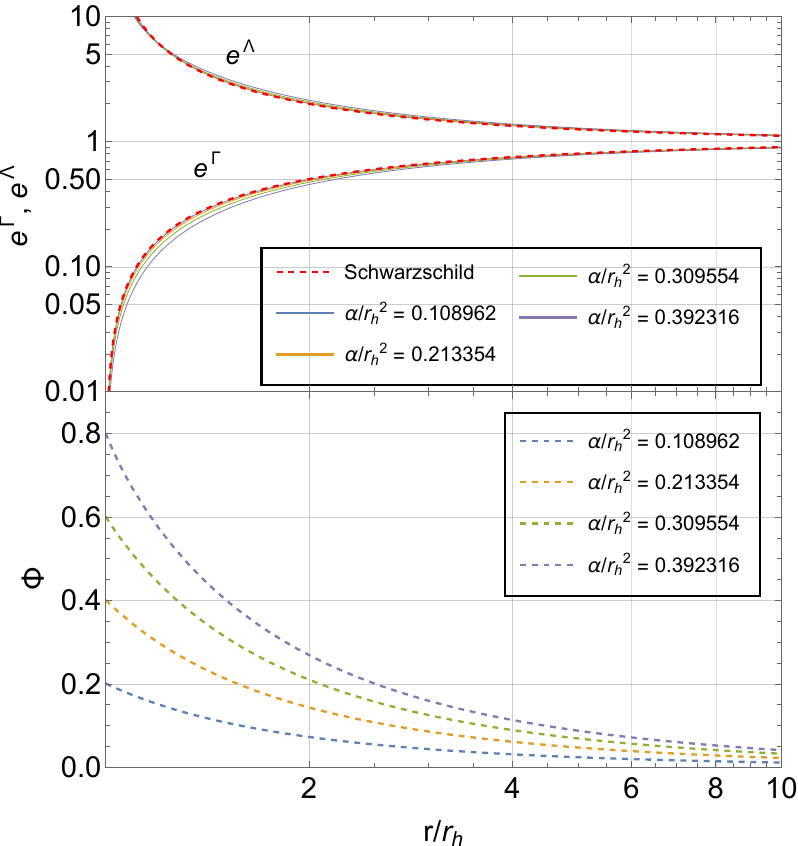}
\caption{(Top) Radial profile of the metric functions $e^{\Gamma(r)}$ and $e^{\Lambda(r)}$ for several EsGB BH solutions with a linear coupling. The Schwarzschild solution corresponds to the red dashed line. (Bottom) Radial profile of the scalar field, $\Phi(r)$, for the same BH solutions.}
\label{fig:metric_linear}
\end{figure}

Fig.~\ref{fig:solutions_linear_phi0_vs_alpha} shows the domain of existence of EsGB BHs with a linear coupling in the ($\Phi_h$, $\talpha$) space. 
These solutions exist along a single branch, developing from the origin until condition~\eqref{eq:inequality} is no longer satisfied, for $|\talpha| = 1/\sqrt{6}$.
Radial profiles of the metric functions and scalar field are shown in Fig.~\ref{fig:metric_linear} for different values of the coupling constant. 
Given that $\talpha$ is restricted to a finite interval, the deviation of the metric functions from their GR counterpart ($\alpha=0$) is bounded from above, with only mild departures in this shift-symmetric case. Moreover, the scalar field profile is bounded from above and below by the solutions for $\Phi(r)$ with $\talpha = \pm 1/\sqrt{6}$.

The most constraining observational bound on the coupling constant in shift-symmetric EsGB theory is 
$\sqrt{\alpha}\leq 1.50\, \text{km}$~\cite{Sanger:2024axs}. Note, however, that this bound relies on post-Newtonian results, and has been challenged in~\cite{Corman:2025wun}.

\subsection{Quadratic coupling}

The quadratic coupling
\be
\label{eq:quad_coupling}
f(\Phi) = \frac{\Phi^2}{4}
\ee
is the simplest choice that allows GR BHs to persist as solutions of EsGB gravity~\cite{Silva:2017uqg}.
Again, the multiplicative factor $1/4$ is purely conventional.

\begin{figure}[t!]
\includegraphics[width=8.3cm]{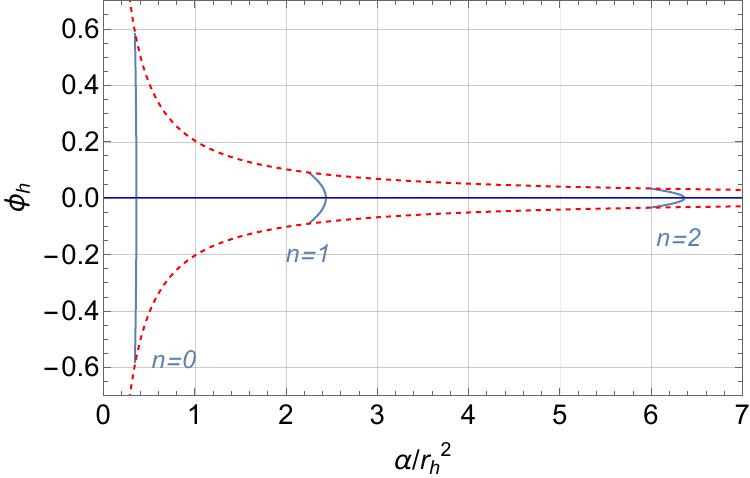}
\caption{First three branches of BH solutions of EsGB with quadratic coupling, $f(\Phi)=\Phi^2/4$, plotted in the ($\Phi_h$, $\alpha/r_h^2$) parameter space. The Schwarzschild solution, represented in blue along the $\Phi_h=0$ axis, is a solution for all $\alpha$, but it becomes unstable at specific values of $\alpha/r_h^2$, leading to the emergence of BHs with scalar hair. These hairy solutions branch out from the $\Phi_h=0$ axis until Eq.~\eqref{eq:inequality} is no longer satisfied. Hairy BHs are restricted to lie within the two red-dashed curves determined by $\Phi_h = \pm(24\alpha^2/r_h^4)^{-1/2}$.
\label{fig:solutions_quadratic_phi0_vs_alpha}}
\end{figure}

The values of $\talpha$ and $\Phi_h$ for which hairy BH solutions exist in this theory are shown in Fig.~\ref{fig:solutions_quadratic_phi0_vs_alpha}, for the fundamental branch ($n=0$) and for the first two overtones ($n=1,2$). As $n$ increases, the parameter range narrows in $\Phi_h$ but widens in $\talpha$. For the $n=0$ solution branch, which introduces the most noticeable changes to Schwarzschild, the coupling parameter is only allowed to vary within $0.346580\lesssim \talpha \lesssim 0.362811$, corresponding to values of the scalar field at the horizon that satisfy $|\Phi_h| \lesssim 0.588967$.
At the scales shown in Fig.~\ref{fig:solutions_quadratic_phi0_vs_alpha}, the $n=0$ branch appears as an almost straight vertical line, but a zoom-in reveals that it bends to the left away from the branching point, similar to the $n=1,2$ branches.

\begin{figure}[t!]
\includegraphics[width=7.5cm]{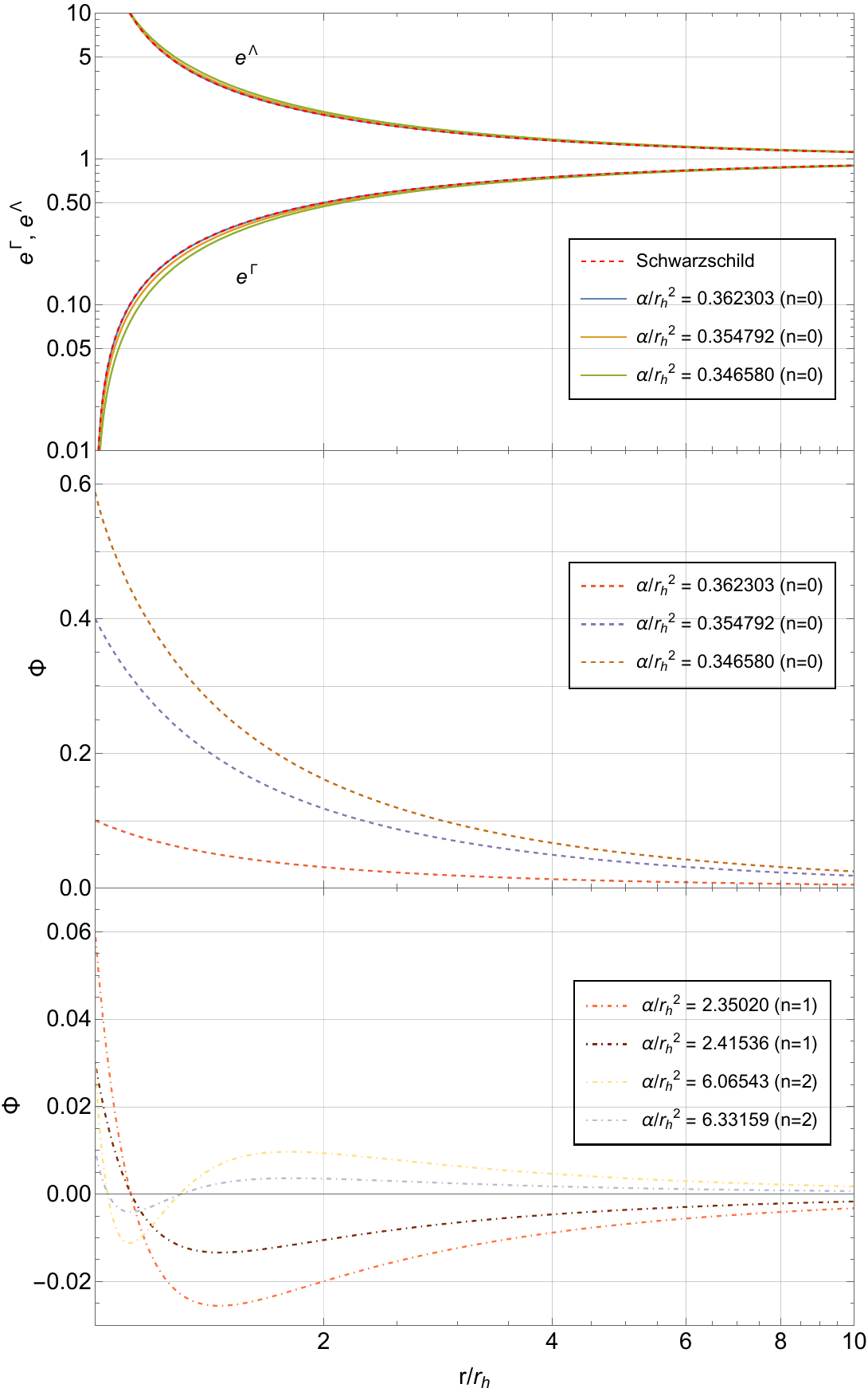}
\caption{(Top) Radial profile of the metric functions $e^{\Gamma(r)}$ and $e^{\Lambda(r)}$ for several $n=0$ EsGB BH solutions with a quadratic coupling. The Schwarzschild solution corresponds to the red-dashed line. (Middle) Radial profile of the scalar field, $\Phi(r)$, for the same $n=0$ BH solutions. (Bottom) Radial profile of the scalar field, $\Phi(r)$, for $n=1$ and $n=2$ EsGB BH solutions with a quadratic coupling.}
\label{fig:metric_quadratic}
\end{figure}

In Fig.~\ref{fig:metric_quadratic}, we show the radial profiles of the metric functions and the scalar field for BH solutions for several values of $\talpha$ and $n$. Note that the metric functions deviate only slightly from Schwarzschild, as the scalar field remains relatively small throughout.
The scalar field profile qualitatively changes as we consider higher $n$, exhibiting an increasing number of nodes. However, its amplitude is correspondingly reduced, which prevents a significant impact on the metric functions. For this reason, we only display the metric functions for the $n=0$ branch of solutions.

It is worth noting that all hairy BHs in EsGB with a purely quadratic coupling function are known to be unstable~\cite{Blazquez-Salcedo:2018jnn, Silva:2018qhn, Minamitsuji:2018xde}. Therefore, we will consider these spacetimes mainly as a benchmark for BHs in EsGB with a special exponential coupling (see next section).

\subsection{Special exponential coupling}

Our last choice of coupling function has a particular exponential form~\cite{Doneva:2017bvd}, 
\be
\label{eq:exp_coupling}
f(\Phi)=\frac{1}{6}(1-e^{-3 \Phi^2 /2})\,.
\ee

\begin{figure}[t!]
\includegraphics[width=7.0cm]{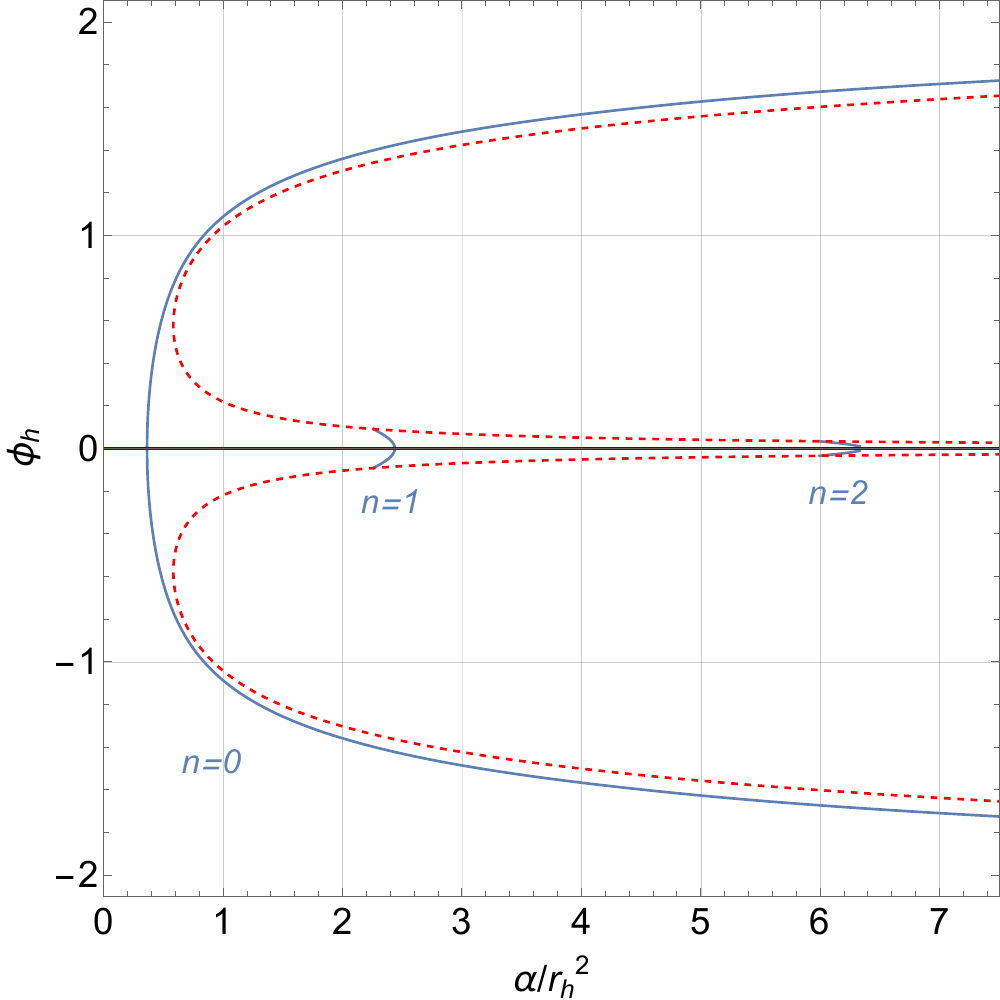}
\caption{First three branches of BH solutions of EsGB with exponential coupling, $f(\Phi)=(1-e^{-3 \Phi^2 /2})/6$, plotted in the ($\Phi_h$, $\alpha/r_h^2$) parameter space. The Schwarzschild solution, represented in blue in the $\Phi_h=0$ axis, is a solution for all coupling parameters $\alpha/r_h^2$. The Schwarzschild solution becomes unstable for $\alpha/r_h^2 > 0.362811$, leading to the emergence of BH solutions with scalar hair. For $n \geq 1$ solutions, these develop in the ($\Phi_h$, $\alpha/r_h^2$)-plane from the $\Phi_h=0$ axis until Eq.~\eqref{eq:inequality} is no longer satisfied. On the other hand, $n=0$ solutions exist for all values of $\Phi_h$. The red, dashed curves were obtained by solving numerically the equality in Eq.~\eqref{eq:inequality}. 
\label{fig:solutions_doneva_phi0_vs_alpha}}
\end{figure}

\begin{figure}[t!]
\includegraphics[width=7.5cm]{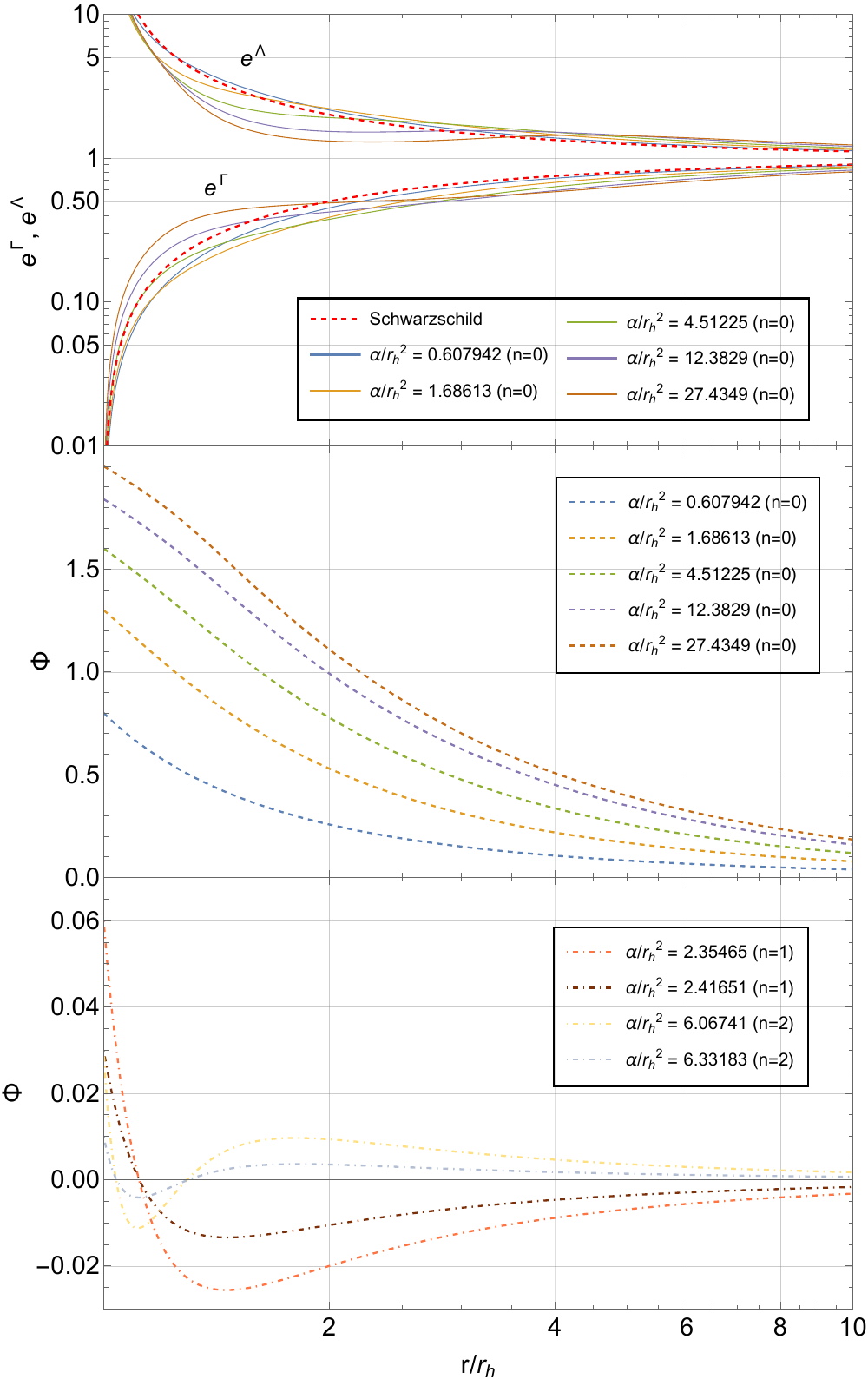}
\caption{(Top) Radial profile of the metric functions $e^{\Gamma(r)}$ and $e^{\Lambda(r)}$ for several $n=0$ EsGB BH solutions with an exponential coupling. The Schwarzschild solution corresponds to the red, dashed line. (Middle) Radial profile of the scalar field, $\Phi(r)$, for the same $n=0$ BH solutions. (Bottom) Radial profile of the scalar field, $\Phi(r)$, for several $n=1$ and $n=2$ EsGB BH solutions with an exponential coupling.}
\label{fig:metric_exponential}
\end{figure}

The BHs in this theory are more interesting than with the quadratic coupling discussed in the previous subsection for two reasons. 
First, some of these hairy BHs are believed to be stable, as we discuss shortly below. 
The second reason is related to Fig.~\ref{fig:solutions_doneva_phi0_vs_alpha}, where we plot the domain of existence of BH solutions in the ($\Phi_h$, $\talpha$) plane. 
Note that, if $|\Phi_h|\ll 1$, we obtain a similar behavior as the one found for the quadratic coupling. For instance, the hairy BHs branch out at exactly the same points. This is expected since $|\Phi_h|\ll 1$ implies $|\Phi(r)|\ll 1$ for all $r>r_h$, in which case the coupling function is well approximated 
by $f(\Phi)=(1-e^{-3 \Phi^2 /2})/6 \simeq \Phi^2/4$. 
For $n\geq1$ solutions, $\Phi_h$ remains bounded to small values, and so the parameters covered in $\talpha$ are very similar to those in the quadratic case. However, for $n=0$ BH solutions, there is no upper bound on $\talpha$ and consequently $\Phi_h$ can take arbitrarily large values. This leads to solutions that significantly differ from Schwarzschild, see Fig.~\ref{fig:metric_exponential}, whereas $n\geq 1$ solutions remain very similar to those obtained in Fig.~\ref{fig:metric_quadratic}. The $n=0$ family of hairy BH solutions branches off from Schwarzschild at a critical value of the coupling constant, given by $\talpha_c = 0.362811$, exactly the same value as in the quadratic coupling case.

It has been shown that all such hairy BHs with $n>0$ are unstable \cite{Blazquez-Salcedo:2018jnn}. On the other hand, the fundamental mode $n=0$ yields a stable hairy BH when the coupling is sufficiently small, specifically $\talpha \lesssim 13.834$.\footnote{This threshold was obtained by converting the bound reported in Refs.~\cite{Blazquez-Salcedo:2018jnn, Blazquez-Salcedo:2020caw, Blazquez-Salcedo:2020rhf}, where it was expressed in units of $M$ instead of $r_h$, and taking into account differences in the definition of the coupling constant.}

Observational constraints on the coupling constant in this specific EsGB theory have been placed by considering GW data from binary BH mergers, strongly disfavoring 
$117\, \text{km} \leq \sqrt{\alpha} \leq 200\, \text{km}$~\cite{Wong:2022wni}.

\section{Black hole mergers in EsGB\label{sec:mergers}}

The event horizon describing the merger between a small BH in EsGB and an infinitely large BH is obtained by computing a particular congruence of null geodesics in the spacetime defined by the small BH in isolation~\cite{Emparan:2016ylg}. The congruence is selected by the condition that the event horizon approaches a null plane at future null infinity. By taking progressive constant time slices, one obtains the time evolution of the merging horizon, as done previously for modified gravity, namely for Einsteinian cubic gravity, in~\cite{Dias:2024wib}.
Determining the location of the merging event horizon by integrating its null generators backwards in time from a late-time stationary state is a robust technique also used occasionally in NR~\cite{Libson:1994dk, Thornburg:2006zb, Masso:1998fi}.

In our analysis, we will take the large BH to be non-hairy. While this is not a matter of concern for the quadratic and exponential couplings considered, for the shift-symmetric theory this assumption requires some explanation, because its BH solutions are necessarily distinct from those of GR. 
However, in the linear case this is justified precisely for infinitely large BHs: when $r_h\to\infty$ the scalar field becomes infinitely suppressed (see Figs.~\ref{fig:solutions_linear_phi0_vs_alpha} and~\ref{fig:metric_linear}).
Indeed, EMR mergers whose primary component is a supermassive BH are prime targets of LISA and thus are expected to be non-hairy.

\subsection{Construction of the merging event horizon}

As just mentioned, the determination of the event horizon hinges upon the calculation of certain null geodesics. 
Let $x^a = (t,r,\theta,\varphi)$ be the coordinates of one such geodesic, parametrized by an affine parameter $\kappa$, and $u^a = dx^a/d\kappa$ denote its tangent velocity vector, in terms of which the geodesic equations are $u^b \nabla_b u^a = 0$.

Given the spherical symmetry of the background BH considered, we may take the geodesic to lie on the equatorial plane without loss of generality, so that $\theta(\kappa)=\pi/2$. 
The three remaining components of the geodesic equations can be written as
\begin{align}
    &\frac{d t}{d\tilde\kappa} = \frac{e^{-\Gamma(r)}}{q} \,, \\
    &\frac{d \varphi}{d\tilde\kappa} = -\frac{1}{r^2} \,, \\
    &e^{\Gamma(r)+\Lambda(r)}\left(\frac{dr}{d\tilde\kappa}\right)^2 + V_{\mathrm{eff}}(r) = \frac{1}{q^2}\label{eq:geodesic_r}\,,
\end{align}
where $\tilde \kappa = \kappa L $ is a rescaled affine parameter and $q = L/E$ is the impact parameter of the geodesic. Here, $E$ and $L$ denote the conserved energy and angular momentum of the geodesic, associated with the Killing vectors $\partial/\partial t$ and $\partial/\partial \varphi$, respectively, whereas $V_{\mathrm{eff}}(r) = e^{\Gamma(r)}/r^2$ represents the effective potential controlling the radial dynamics.

After specifying the function $f(\Phi)$ and the coupling parameter $\talpha$, the metric functions $\Gamma(r)$ and $\Lambda(r)$ can be obtained using the procedure from Section~\ref{sec:EsGB}, and the geodesic equations can be solved for each impact parameter $q$. 
To this end, it is useful to parametrize the geodesics with the radial coordinate $r$ instead of the affine parameter $\tilde \kappa$. With this substitution, the geodesic equations for the $t$ and $\varphi$ coordinates become
\be
\frac{d t}{d r} = \pm \frac{e^{\frac{\Lambda(r)-\Gamma(r)}{2}}}{q \sqrt{\frac{1}{q^2} - V_{\mathrm{eff}}(r)}} \,, \qquad
\frac{d \varphi}{d r} = \mp \frac{e^{\Gamma(r)+\Lambda(r)}}{r^2 \sqrt{\frac{1}{q^2} -  V_{\mathrm{eff}}(r)}}  \,.
\label{eq:PHIdotTdot}
\ee
The radial coordinate that maximizes $V_\mathrm{eff}$ is the photon sphere radius $r_{\mathrm{ph}}$ of the small BH.
For sufficiently small impact parameters $q$, such that $1/q^2 > V_{\mathrm{eff}}(r_{\mathrm{ph}}) \equiv 1/{q^2_{\mathrm{ph}}}$, the denominators in Eq.~\eqref{eq:PHIdotTdot} never vanish and the integration can be performed readily. This situation corresponds to generators that do not bounce off the potential barrier of the small BH. As a consequence such generators either end up at $r=r_h$, at the infinite past, or they intersect the caustic line\footnote{Caustic points correspond to events at which multiple generators cross and enter the event horizon~\cite{Gadioux:2023pmw}.} $\varphi = \pi$ at some finite time.
In the language of \cite{Emparan:2016ylg} (see Fig. 5 therein), the former have $q\leq q_c$ whereas the latter have $q_c<q<q_{\mathrm{ph}}$. Here, $q_c$ is the critical impact parameter separating non-caustic generators from caustic generators.

Generators with impact parameters $q \geq q_{\mathrm{ph}}$ can be analyzed similarly. The characteristic novel feature of these geodesics is the presence of a turning point at some radius of closest approach to the small BH. There are two cases to be distinguished: either the turning point occurs before reaching the caustic line (i.e., $|\varphi|<\pi$), or it occurs afterwards (and will, therefore, not be part of the event horizon). The threshold separating these two classes of generators happens at some impact parameter $q_*$, determined by having its turning point exactly on the caustic line, at some radial coordinate $r_*$.

In summary, we can divide generators into two different categories: non-caustic generators, with impact parameters $q \leq q_c$, and caustic generators, with $q>q_c$. The latter can further be subdivided based on whether $q_c< q < q_*$ or $q>q_*$, corresponding to whether they enter the caustic from the side of the small BH or from the side of the large BH, respectively.

An important conclusion of the above discussion is that 
\be
    q_c < q_{\mathrm{ph}} < q_*\,.
\ee
Since $q_{\mathrm{ph}}$ is a local quantity, it is easier to compute than $q_c$ or $q_*$, and can therefore be used to interpret the results of the latter.

To solve the geodesic equations~\eqref{eq:PHIdotTdot}, we must provide boundary conditions that ensure that the congruence of null geodesics approaches a null plane at future null infinity. This is done by expanding the geodesic equations for large $r$, in which case the functions $t(r)$ and $\varphi(r)$ can be straightforwardly integrated to give~\cite{Emparan:2016ylg}
\bea
\varphi_q(r\to\infty) \!\!&=&\!\! \varphi_{\infty} + \frac{q}{r} + O(r^{-3})\,, \\
t_q(r\to\infty) \!\!&=&\!\! t_{\infty}+r + 2M \ln(r/2M) + O(r^{-1})\,,
\eea
where we have used the asymptotic behavior of $e^{\Gamma(r)}$ and $e^{\Lambda (r)}$. To satisfy the boundary conditions, the constants $\varphi_{\infty}$ and $t_{\infty}$ must be independent of $q$ so that the generators are asymptotically parallel to each others and all reach future null infinity at the same retarded time. Since the spacetime of the small BH is static and spherically symmetric, we can choose $\varphi_{\infty}=t_{\infty}=0$ without loss of generality.

A final technical comment is in order regarding the integration of the geodesic equations. As explained above, generators with impact parameter $q>q_*$ reach a turning point before hitting the caustic line at $\varphi = \pi$. In these cases, Eq.~\eqref{eq:PHIdotTdot} becomes singular, which is just a result of having parametrized the geodesics with the radial coordinate instead of the affine parameter $\tilde\kappa$. 
This issue can be circumvented by exploiting the spherical symmetry of the small BH. Indeed, since the geodesics must exhibit symmetry about their turning point, we can continue the integration by a simple reflection that ensures continuity of the functions $t$ and $\varphi$ and of their derivatives at the turning point.

\subsection{Merger characterization}

As noted in the previous section, caustic generators have impact parameters $q>q_c$ and enter the caustic line $\varphi = \pi$ at finite times $t$.  
The generators with impact parameter $q=q_*$
determine the event where the two individual pre-merger horizons pinch on. At this event, the turning point $r=r_*$ lies exactly at the caustic line and determines the maximum radial distortion of the small BH due to the gravitational pull of the large BH, whereas the time $t=t(r_*)\equiv t_*$ defines the instant at which the pinch occurs. Thus, numerically, we can determine $q_*$, $r_*$ and $t_*$ by solving the system:
\be
    V_\mathrm{eff}(r_*)=1/q_*^2 \,, \qquad \varphi(r_*)=\pi \,, \qquad t(r_*)=t_*\,.
\ee

Using this, one can introduce two other quantities~\cite{Emparan:2016ylg} which are central to our analysis: 
(1) the duration of the merger, $\Delta_*$, is defined as the retarded time interval between the pinch-on instant $t_*$ and the asymptotic retarded time of the central generator with $q=0$; 
(2) the increase in area of the small BH during the merger can be computed by noting that, at early times, the small BH horizon has an area $\mathcal{A}_{in} = 4\pi r_h^2$, whereas at late times this area has grown by an amount $\Delta \mathcal{A}_{\mathrm{smallBH}}=\left[q_*^2/(4r_h^2)-1\right]\mathcal{A}_{in}$. The additional area is due to the contribution of those generators that enter the caustic from the side of the small BH, yielding a final area given by the area of a disk with radius $q_*$.

This analysis allows us to quantify the impact of the coupling function $f(\Phi)$ on the defining characteristics of a horizon-fusing process, and to scan the whole $\talpha$ parameter space. Below, we study the BH merger properties in EsGB for the three models described in Section~\ref{sec:EsGB}.

\subsubsection{Linear coupling\label{subsubsec:linear}}

Recall that, for the linear coupling function $f(\Phi)=\Phi/4$, the BH solutions are necessarily hairy if $\talpha \neq 0$. Additionally, they all belong to a single branch that connects to the GR black holes at $\talpha = 0$, in contrast to those from theories that admit spontaneously scalarized BHs.

\begin{figure}[t!]
\includegraphics[width=8.3cm]{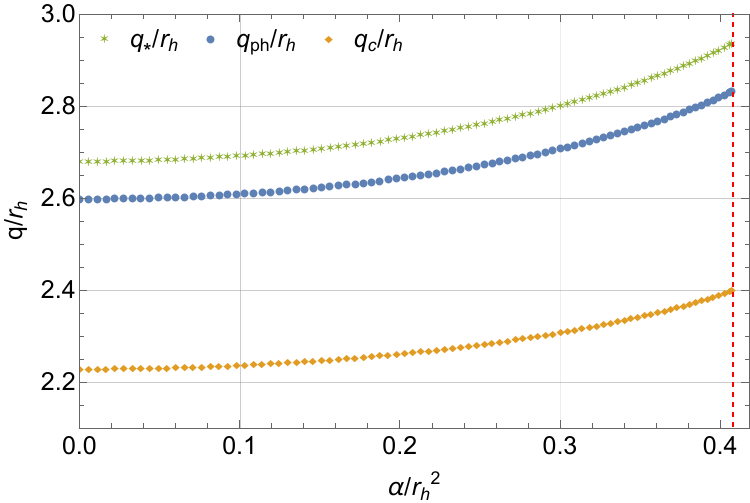}
\caption{Critical impact parameters, $q_c/r_h$ and $q_*/r_h$, as a function of the coupling constant $\alpha/r_h^2$, for the merger of an EsGB BH with a linear coupling function, $f(\Phi)=\Phi/4$. Also shown are the impact parameter related to the maximum of $V_{\mathrm{eff}}$, $q_{\mathrm{ph}}$, as well as the line marking the upper threshold of $\alpha/r_h^2$ for which BH solutions exist, $\alpha/r_h^2 <1/\sqrt{6}$.\label{fig:impact_parameters_linear}}
\end{figure}

We plot the impact parameters $q_\mathrm{ph}$, $q_c$, and $q_*$, introduced before, against the coupling parameter $\talpha$ in Fig.~\ref{fig:impact_parameters_linear}. It is manifest that $q_c < q_{\mathrm{ph}}<q_*$, as expected from the discussion above. Additionally, all three functions exhibit a similar monotonic growth with increasing $\talpha$.

\begin{figure*}[t!]
    \centering
    \begin{minipage}{0.33\linewidth}
        \centering
        \includegraphics[width=\linewidth]{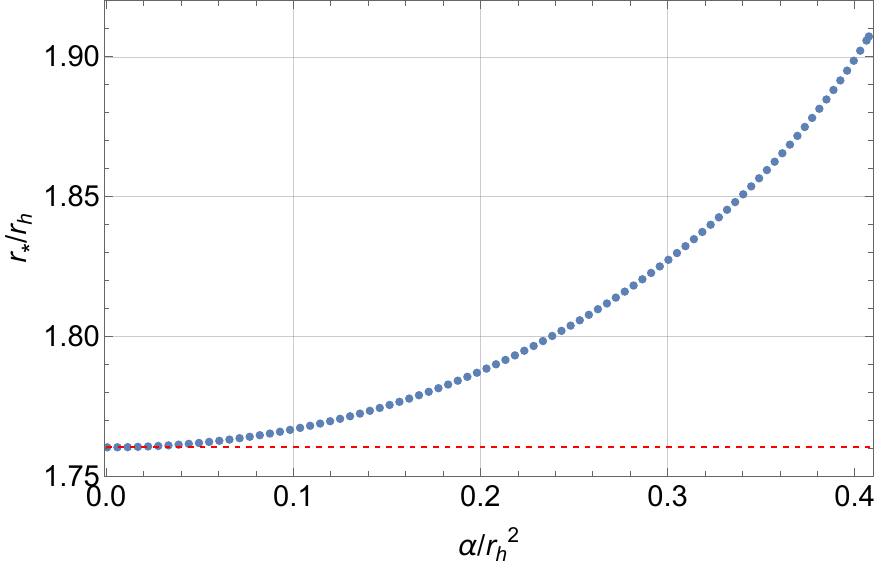}
    \end{minipage}\hfill
    \begin{minipage}{0.33\linewidth}
        \centering
        \includegraphics[width=\linewidth]{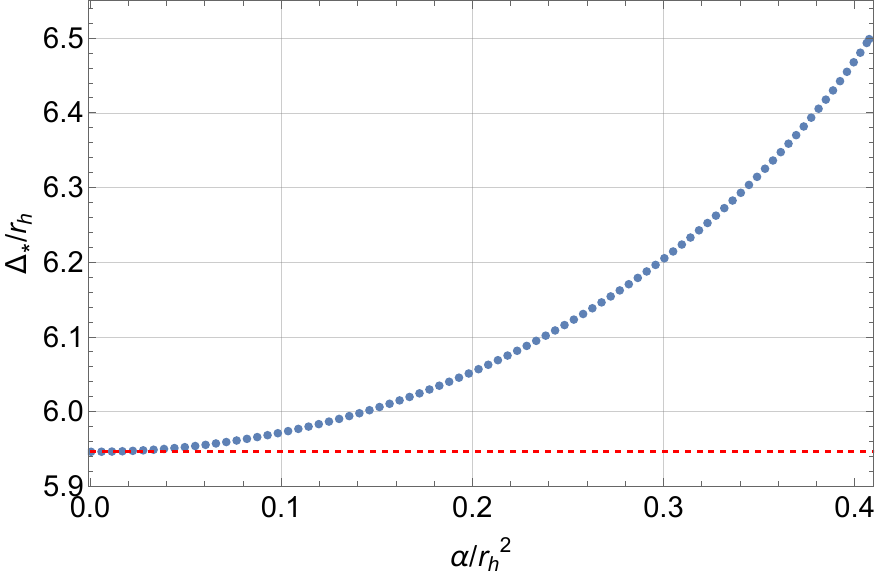}
    \end{minipage}
    \begin{minipage}{0.33\linewidth}
        \centering
        \includegraphics[width=\linewidth]{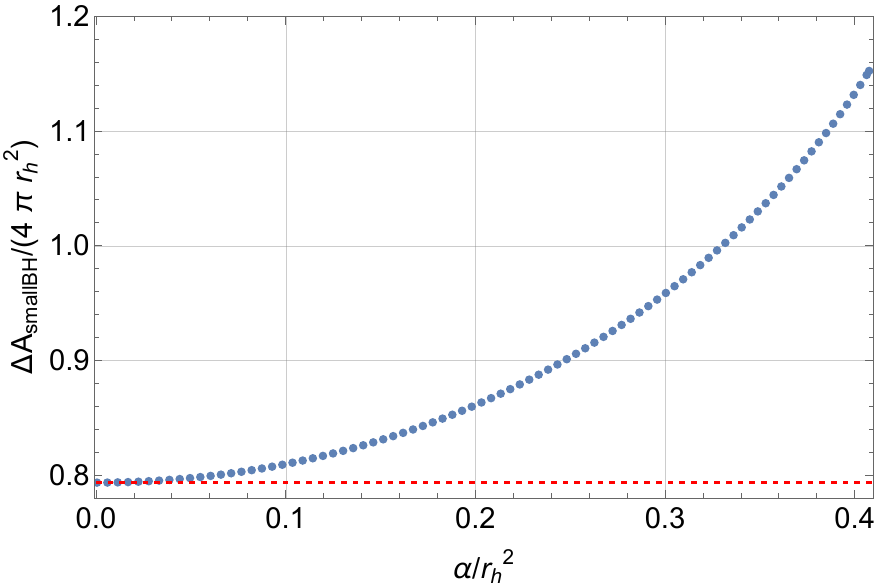}
    \end{minipage}
    \caption{(Left) Maximum distortion of the small BH, $r_*$; (middle) duration of the merger, $\Delta_*$; and (right) increase in the small BH area, $\mathcal{A}_{\mathrm{smallBH}}$, as functions of the coupling constant, $\alpha/r_h^2$, for EsGB solutions with linear coupling, $f(\Phi)=\Phi/4$. The dotted red lines correspond to the values obtained for the Schwarzschild solution in \cite{Emparan:2016ipc}, namely $r_*=1.76031\, r_h$, $\Delta_*=5.94676\, r_h$, and $\mathcal{A}_{\mathrm{smallBH}} = 0.79356 \times (4\pi r_h^2)$.}\label{fig:overall_linear}
\end{figure*}

Figure~\ref{fig:overall_linear} shows the maximal radial distortion, the duration of the merger, and the increase in the small BH area, defined in the beginning of this section, as functions of $\talpha$.
It is manifest that larger coupling constants yield longer merging phases. The two other quantities follow a similar trend, with the small BH undergoing a larger radial distortion and experiencing a larger increase in area.

It is interesting to compare results for the merger duration to those obtained using NR in~\cite{Corman:2025wun}. However, such a comparison is challenging because that study considered comparable mass binaries, while we work in the extreme mass ratio regime. More importantly, the strict EMR limit does not allow us to normalize physical quantities with respect to the total mass of the binary system, which is the normalization adopted in~\cite{Corman:2025wun}. We attempt to make a sensible comparison in Section~\ref{sec:comparison}.

The coupling and mass of the small BH chosen in \cite{Corman:2025wun}, 
corresponds to $\talpha=0.145168$.
This gives $r_*/r_h=1.77417$, $\Delta_*/r_h=6.00111$, and $\Delta A_\mathrm{smallBH}/(4 \pi r_h^2) = 0.828108$, which, relative to the Schwarzschild values. represent increments of $0.79 \%$, $0.91 \%$, and $4.35 \%$, respectively.

\subsubsection{Quadratic coupling\label{subsubsec:quadratic}}

We now move on to the case of a quadratic coupling function,  $f(\Phi)=\Phi^2/4$. Since all the BH solutions arising from a purely quadratic coupling are unstable~\cite{Blazquez-Salcedo:2018jnn, Silva:2018qhn, Minamitsuji:2018xde}, these results mainly serve to connect with those for the special exponential coupling. As discussed in Section~\ref{sec:EsGB}, the parameter space ($\Phi_h$, $\talpha$) for which BHs exist is rather small, see Fig.~\ref{fig:solutions_quadratic_phi0_vs_alpha}. 
Additionally, in contrast to the linear coupling theory, decreasing the coupling parameter increases the value of the scalar field at the horizon for all solution branches.

\begin{figure}[t!]
\includegraphics[width=8.3cm]{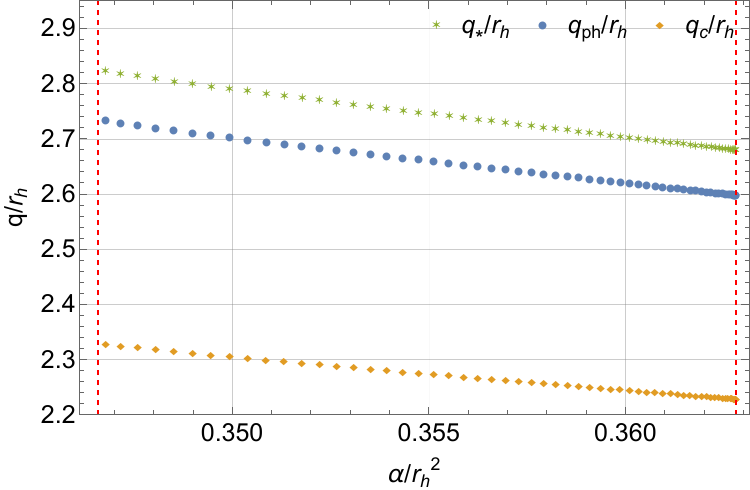}
\caption{Critical impact parameters, $q_c/r_h$ and $q_*/r_h$, as a function of the coupling parameter $\alpha/r_h^2$, for the merger of an $n=0$ branch EsGB BH with a quadratic coupling function, $f(\Phi)=\Phi^2/4$. Also shown are the impact parameter related to the maximum of $V_{\mathrm{eff}}$, $q_{\mathrm{eff}}$, as well as the lines marking the lower and upper thresholds of $\alpha/r_h^2$ for which BH solutions exist, $0.346580\lesssim \alpha/r_h^2 \lesssim 0.362811$.\label{fig:impact_parameters_quadratic}}
\end{figure}

Proceeding as for the linear case, 
Fig.~\ref{fig:impact_parameters_quadratic} shows how the impact parameters $q_\mathrm{ph}$, $q_c$, and $q_*$ depend on the coupling constant $\talpha$ for the $n=0$ branch solutions. All the impact parameters decrease monotonically with $\talpha$, and consistently with our previous discussion, $q_c < q_{\mathrm{ph}}<q_*$, until they merge with the corresponding Schwarzschild values at $\talpha=\talpha_c=0.362811$.

\begin{figure*}[t!]
    \centering
    \begin{minipage}{0.33\linewidth}
        \centering
        \includegraphics[width=\linewidth]{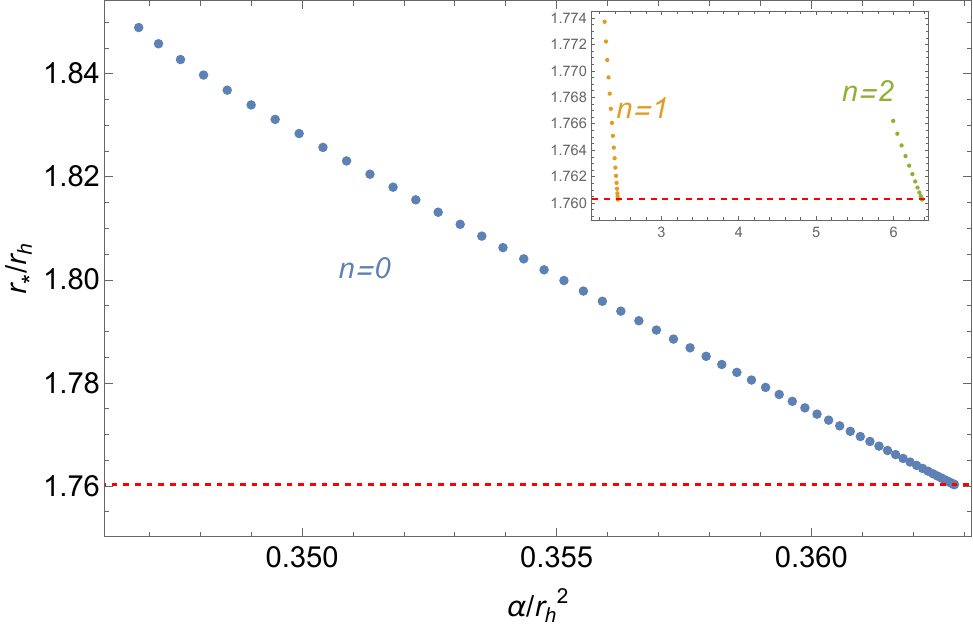}
    \end{minipage}\hfill
    \begin{minipage}{0.33\linewidth}
        \centering
        \includegraphics[width=\linewidth]{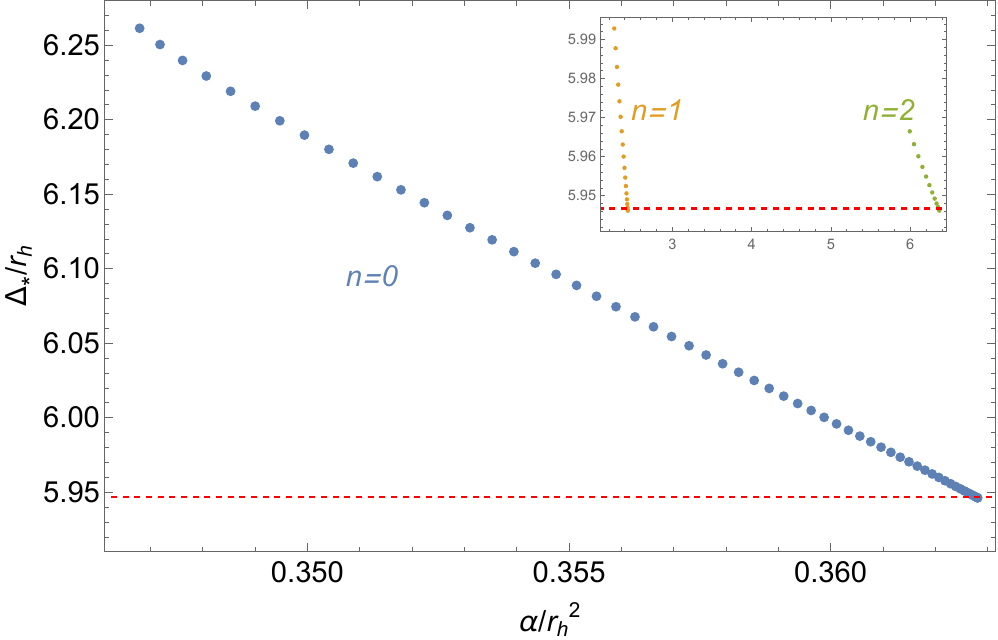}
    \end{minipage}
    \begin{minipage}{0.33\linewidth}
        \centering
        \includegraphics[width=\linewidth]{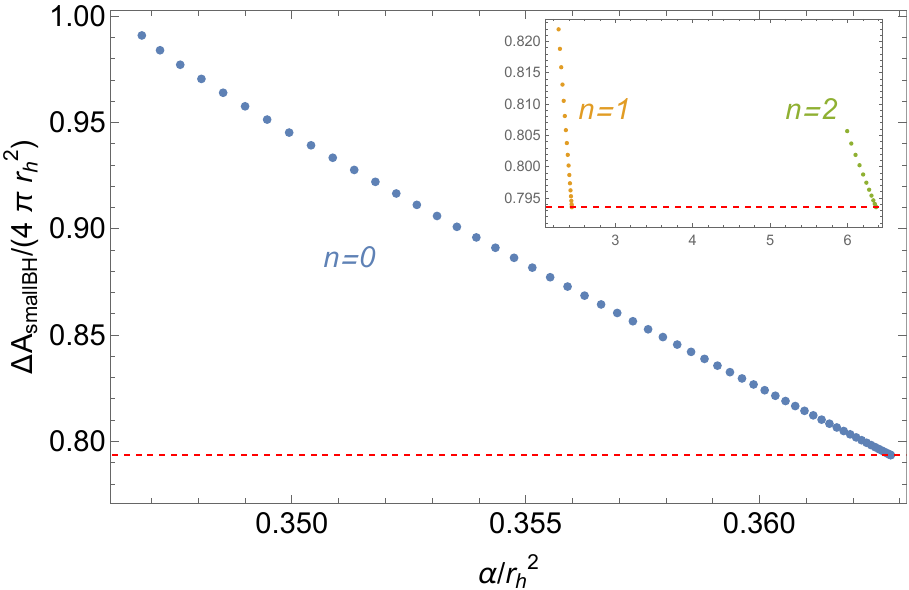}
    \end{minipage}
    \caption{(Left) Maximum distortion of the small BH, $r_*$; (middle) duration of the merger, $\Delta_*$; and (right) increase in the small BH area, $\mathcal{A}_{\mathrm{smallBH}}$, as functions of the coupling parameter, $\alpha/r_h^2$, for $n=0$ (blue), $n=1$ (yellow) and $n=2$ (green) EsGB solutions with quadratic coupling, $f(\Phi)=\Phi^2/4$. The dotted red lines correspond to the values obtained for the Schwarzschild solution in \cite{Emparan:2016ipc}, namely $r_*=1.76031 r_h$, $\Delta_*=5.94676 r_h$, and $\mathcal{A}_{\mathrm{smallBH}} = 0.79356 \times (4\pi r_h^2)$.}\label{fig:overall_quadratic}
\end{figure*}

The characteristic quantities of the merger introduced in the beginning of this section are plotted as a function of $\talpha$ in Fig.~\ref{fig:overall_quadratic}, for $n=0, 1, 2$. We note that increasing $\talpha$ results in faster mergers, with the small BH undergoing less distortion and experiencing a smaller increase in area. This is consistent with the results of Fig.~\ref{fig:impact_parameters_quadratic}. Note that, in this case, decreasing the coupling constant increases the value of the scalar field at the horizon for all solution branches (see Fig.~\ref{fig:solutions_quadratic_phi0_vs_alpha}). This explains why the merger properties follow the opposite trend compared to the linear coupling theory.

Additionally, the rate of decrease in these quantities is steeper for lower branches, which translates into the fact that $\Phi_h$ changes faster with $\talpha$ for lower-branch solutions than for higher-branch ones, see Fig.~\ref{fig:solutions_quadratic_phi0_vs_alpha}. It is worth mentioning that in the $\Phi_h \rightarrow 0$ limit --- in which case $\talpha$ is the maximum value allowed for each branch --- we recover the Schwarzschild results, obtained in \cite{Emparan:2016ylg}.

\subsubsection{Special exponential coupling\label{subsubsec:exponential}}

\begin{figure}[t!]
\includegraphics[width=8.3cm]{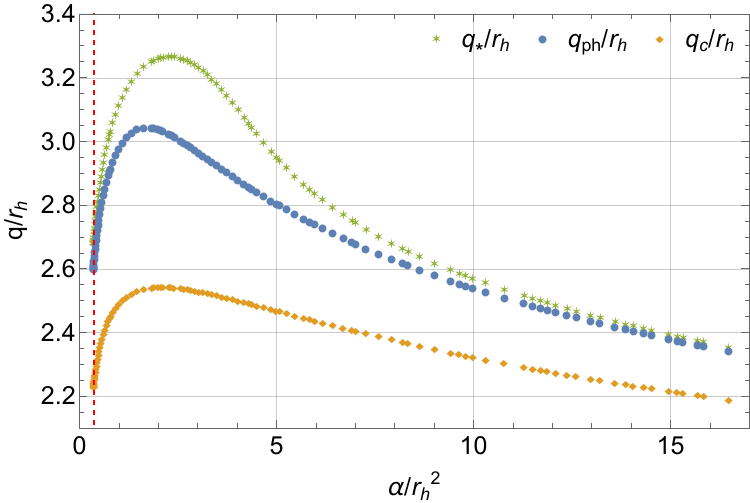}
\caption{Critical impact parameters, $q_c/r_h$ and $q_*/r_h$, as a function of the coupling parameter $\alpha/r_h^2$, for the merger of an $n=0$-branch EsGB BH with an exponential coupling function, $f(\Phi)=(1-e^{-3\Phi^2/2})/6$. Also shown are the impact parameter related to the maximum of $V_{\mathrm{eff}}$, $q_{\mathrm{eff}}$, as well as the line marking the lower threshold of $\alpha/r_h^2$ for which BH solutions exist, $\alpha/r_h^2 \gtrsim 0.362811$.\label{fig:impact_parameters_Doneva}}
\end{figure}

Here we perform the same analysis as before, but for the special exponential coupling function, $f(\Phi)=(1-e^{-3\Phi^2/2})/6$, which also admits spontaneously scalarized BH solutions.
As before, we plot the impact parameters $q_\mathrm{ph}$, $q_c$ and $q_*$ in Fig.~\ref{fig:impact_parameters_Doneva} as a function of $\talpha$, for $n=0$ branch solutions.
These impact parameters show similar qualitative behavior, and satisfy the ordering $q_c < q_{\mathrm{ph}}<q_*$, in agreement with our previous discussion. Moreover, we observe that $q_*$ asymptotes $q_{\mathrm{ph}}$ for large $\talpha$.

\begin{figure*}[t!]
    \centering
    \begin{minipage}{0.33\linewidth}
        \centering
        \includegraphics[width=\linewidth]{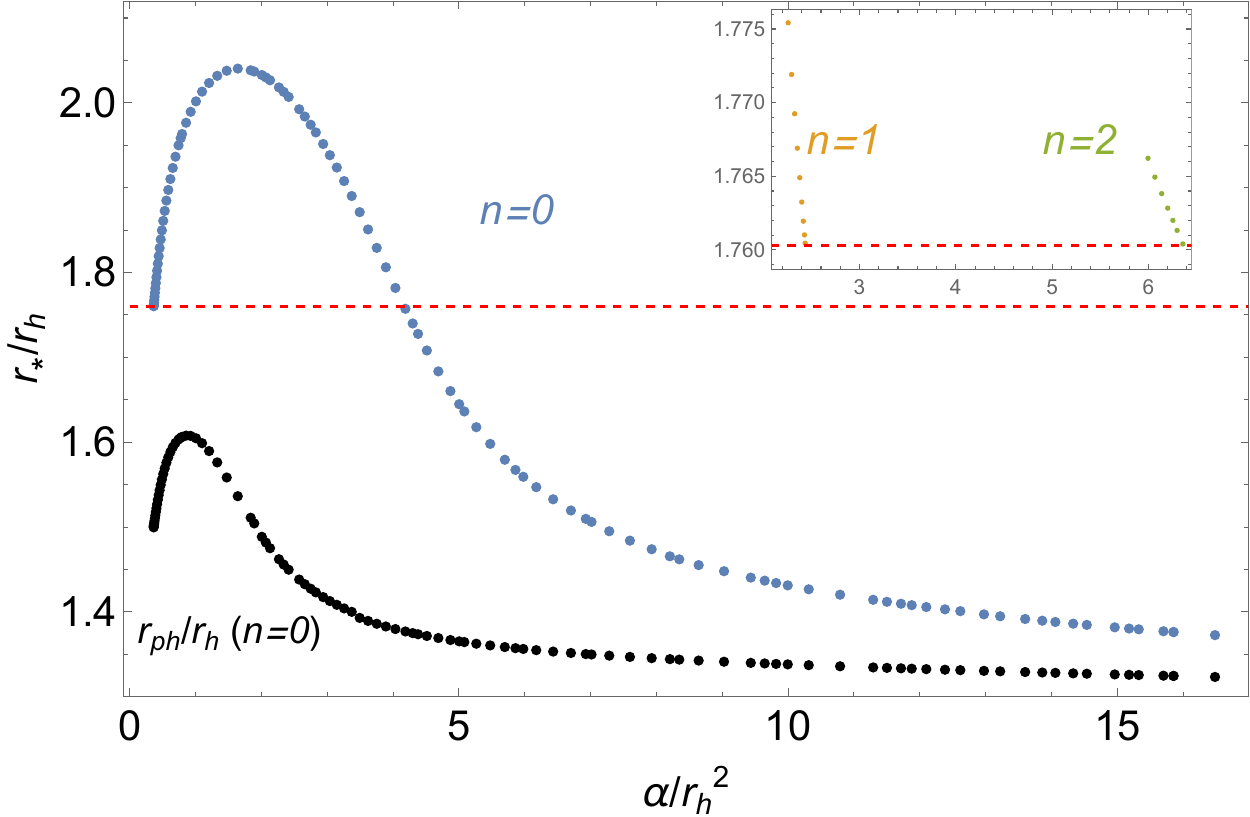}
    \end{minipage}\hfill
    \begin{minipage}{0.33\linewidth}
        \centering
        \includegraphics[width=\linewidth]{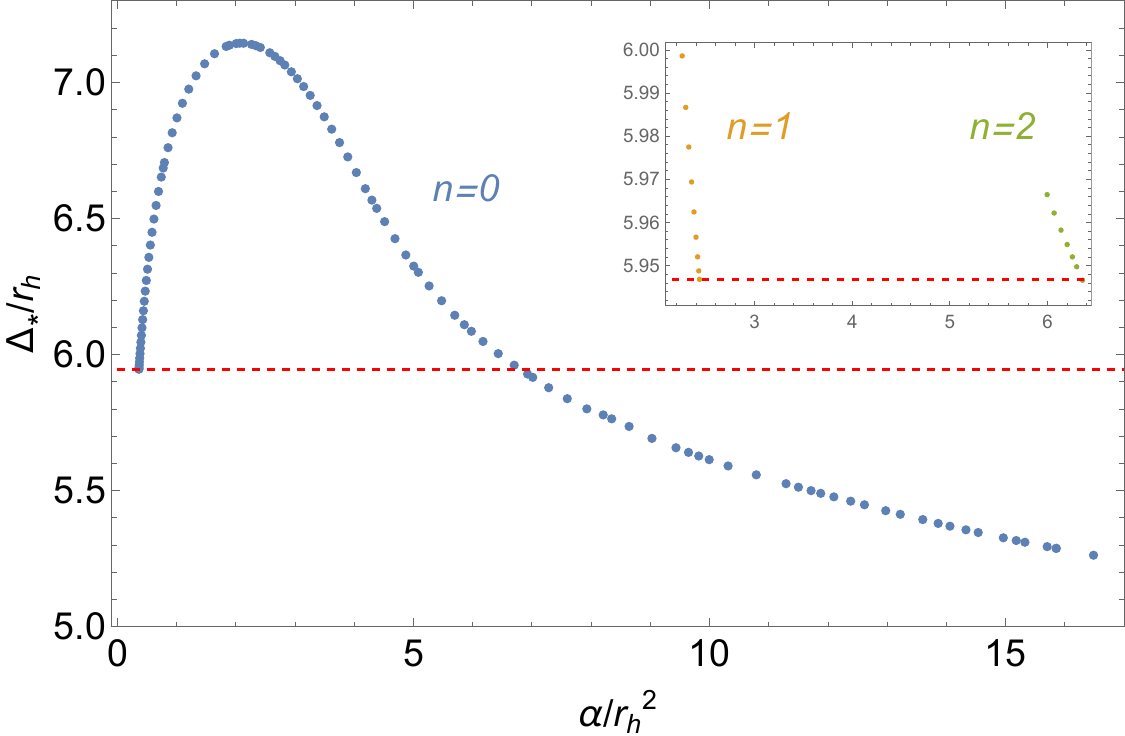}
    \end{minipage}
    \begin{minipage}{0.33\linewidth}
        \centering
        \includegraphics[width=\linewidth]{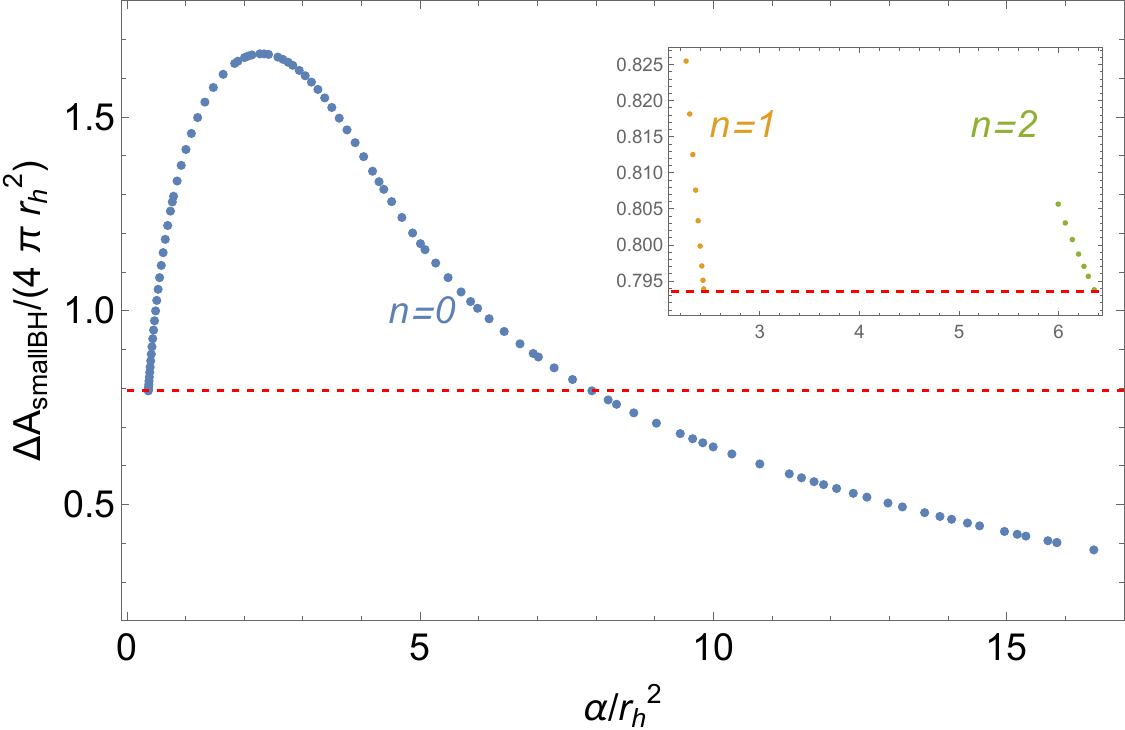}
    \end{minipage}
    \caption{(Left) Maximum distortion of the small BH, $r_*$; (middle) duration of the merger, $\Delta_*$; and (right) increase in the small BH area, $\mathcal{A}_{\mathrm{smallBH}}$, as functions of the coupling parameter, $\alpha/r_h^2$, for $n=0$ (blue), $n=1$ (yellow) and $n=2$ (green) EsGB solutions with exponential coupling, $f(\Phi)=(1-e^{-3 \Phi^2 /2})/6$. The dotted red lines correspond to the values obtained for the Schwarzschild solution in \cite{Emparan:2016ipc}, namely $r_*=1.76031 r_h$, $\Delta_*=5.94676 r_h$, and $\mathcal{A}_{\mathrm{smallBH}} = 0.79356 \times (4\pi r_h^2)$.}\label{fig:overall_Doneva}
\end{figure*}

Fig.~\ref{fig:overall_Doneva} shows how the maximal radial distortion of the small BH, the merger duration, and the area increment depend on $\talpha$, for $n=0,1,2$ solutions. 
An interesting feature to note is that, for $n=0$ solutions, all these quantities attain a maximum at a finite value of $\talpha$.
Starting from the Schwarzschild value at $\talpha = \talpha_c = 0.362811$, they increase with $\talpha$ and reach their respective maxima at distinct values: at $\talpha = 1.66034$ for $r_*/r_h$; $\talpha = 2.09473$ for $\Delta_*/r_h$; and $\talpha = 2.29254$ for $A_\mathrm{smallBH}/(4 \pi r_h^2)$. This mismatch highlights the nonlinearities of the merging process. For larger $\talpha$, they decrease and eventually drop below their corresponding Schwarzschild values. Nevertheless, we observe that the maximum of $r_*$ tracks very closely the maximum of $q_{\mathrm{ph}}$. Similarly, the maximum of $\Delta_*$ closely follows the maximum of $q_c$. And from the definition of $\mathcal{A}_{\mathrm{smallBH}}$ it is clear that its maximum occurs at the same value of the coupling constant that maximizes $q_*$. 

The non-monotonicity displayed in all the quantities analyzed can be attributed to a similar behavior of the photon sphere radius of an isolated hairy BH, $r_{\mathrm{ph}}$, which is also plotted in Fig.~\ref{fig:overall_Doneva}. Consistently with the behaviors of $q_*$ and $q_{\mathrm{ph}}$, we note that $r_*$ appears to asymptote $r_{\mathrm{ph}}$ for large $\talpha$.

\section{Comparison with numerical simulations\label{sec:comparison}}

In all plots displayed up to now, we have represented all length (and time) scales in units of the horizon radius of the small BH, $r_h$.
In order to connect with recent numerical simulations of hairy BH mergers~\cite{Corman:2025wun,Capuano:2026lhs} it is desirable to express our results in units of the total initial ADM mass, $\ADM$, which is the convention usually adopted in NR studies.

In this section we perform a conversion of our results to make a crude comparison with NR simulations. In performing such a comparison, one should keep in mind that Refs.~\cite{Corman:2025wun,Capuano:2026lhs} evolved inspiralling comparable-mass binaries with a finite total initial mass, whereas we considered head-on EMR mergers with an infinite total mass. Moreover, our approach does not account for the emission of gravitational and scalar radiation in the process. These differences imply that a precise comparison is out of reach at present. Nevertheless, we attempt a qualitative comparison in the regime where the small hairy BH is perturbatively close to the Schwarzschild solution of GR.
Since the coupling constant $\alpha$ has dimensions of $(\text{length})^2$, we also need to convert our results from $\talpha$ to
\be
\halpha \equiv \frac{\alpha}{\ADM^2}\,.
\label{eq:alphahat_def}
\ee

First, let us focus on the linear coupling case, where $f(\Phi)=\Phi/4$. Our results (see Fig.~\ref{fig:overall_linear}) indicate that, for small $\talpha$, $\Delta_*/r_h$ grows quadratically,
\be
\label{eq:DeltarhSeries}
\frac{\Delta _*}{r_h} = 5.94676 + 2.4\, \talpha^2 + O(\talpha^3)\,.
\ee
Our aim is to obtain $\Delta_*/\ADM$ from this, as a function of $\halpha$.
Such a relation is non-trivial, and this is where working perturbatively in $\alpha$ becomes advantageous. 
It is important to keep in mind that $\ADM$ is distinct from the mass $M$ of the small BH, which was used in all previous sections. These two masses are related through $\ADM = M_\infty + M$, where $M_\infty$ is the (formally infinite) ADM mass of the large BH, and was argued to be $\alpha$-independent in Section~\ref{sec:mergers}. Clearly, $\ADM$ is also formally infinite, but we are mainly interested in its dependence on $\alpha$, which is entirely dictated by the mass of the small BH as a function of $\alpha$.
Further details are provided in Appendix~\ref{app:details_comp}.

We start by noting that $M$ is related to the coupling constant $\talpha$ through~\cite{Yunes:2011we, Sotiriou:2014pfa}\footnote{The difference in the numerical coefficient in front of $\talpha^2$ as compared to~\cite{Sotiriou:2014pfa} is only due to the distinct normalization of the terms in the action~\eqref{eq:Action}. Upon making the necessary conversions, our results match precisely.}
\begin{equation}
\label{eq:MassAlpha}
    \frac{M}{r_h} = \frac{1}{2} + \frac{49}{160} \talpha^2 + O(\talpha^3)\,.
\end{equation}
Here we used the fact that for GR one has $M(0)/r_h=1/2$.
This perturbative expansion is in excellent agreement with our numerical results shown in the left panel of Fig.~\ref{fig:mass_vs_alpha} in Appendix~\ref{app:details_comp}.
Given the simple relation between $M$ and $\ADM$, we have
\be
\label{eq:FinalMassModGrav}
    \frac{\ADM}{r_h} = b + c\, \talpha^2 + O(\talpha^3)\,,
\ee
with $b>0$ and $c>0$ being $\alpha$-independent quantities. For further details about the derivation, including estimates for the gravitational and scalar radiation, see Appendix~\ref{app:details_comp}. There it is also shown that it is appropriate to neglect the energy emitted in both channels for the purposes of this section, and in the EMR regime.

The relation~\eqref{eq:FinalMassModGrav} allows one to translate the local behavior of any function of $\halpha$ into the corresponding local behavior in terms of $\talpha$ and vice versa, using the chain rule of differentiation. It is also necessary for relating the behaviors of $\Delta_*/\ADM$ and $\Delta_*/r_h$. Since the coefficient $b$ in~\eqref{eq:FinalMassModGrav} is formally infinite, it is only possible to describe how the dimensionless combination $\Delta_*/\ADM$ depends on the coupling $\halpha$ in terms of a formal series expansion in the coupling constant $\halpha$,
\be
\label{eq:DeltaMtotSeries}
\frac{\Delta_*}{\ADM} 
= \left.\frac{\Delta_*}{\ADM}\right|_{\halpha=0}
+
\frac{\halpha^2}{2} \frac{d^2}{d\halpha^2} \left[ \frac{\Delta_*}{\ADM} \right]_{\halpha=0} 
+ O(\halpha^3)\,.
\ee

In Appendix~\ref{app:details_comp} we show that the coefficient of the linear term in this expansion indeed vanishes, and that the coefficient of the quadratic term is strictly positive. 
Thus, our results from Section~\ref{subsubsec:linear} predict that $\Delta_*/\ADM$ increases as a function of $\alpha/\ADM^2$, at least for small coupling constant, in line with the findings of~\cite{Corman:2025wun}.

Now we repeat the preceding analysis for the case of the special exponential coupling function~\eqref{eq:exp_coupling}, which is the general type of scalar-Gauss-Bonnet coupling adopted in Ref.~\cite{Capuano:2026lhs}. Note, however, that none of the simulations performed there considered a coupling function that matches expression~\eqref{eq:exp_coupling} exactly.

We relegate details of this analysis to Appendix~\ref{app:details_comp_exp}, highlighting here only the main results.
First, Fig.~\ref{fig:overall_Doneva} indicates that, for $\talpha\simeq\talpha_c = 0.362811$, $\Delta_*/r_h$ grows linearly,
\be
\label{eq:DeltarhSeries_exp}
\frac{\Delta _*}{r_h} = 5.94676 + 3.0(\talpha-\talpha_c) + O((\talpha-\talpha_c)^2)\,.
\ee
Furthermore, in the special exponential case we have the following relation between $\ADM$ and $\talpha$:
\be
\label{eq:FinalMassModGrav_exp}
    \frac{\ADM}{r_h} = \beta + \gamma\, (\talpha - \talpha_c) + O((\talpha - \talpha_c)^2)\,,
\ee
where $\beta$ and $\gamma$ are reported in Eq.~\eqref{eq:value_gamma}.
Using Eqs.~\eqref{eq:DeltarhSeries_exp} and~\eqref{eq:FinalMassModGrav_exp}, and once again converting from $\talpha$ to $\halpha$, we obtain
\bea
\frac{\Delta_*}{\ADM} 
&=& \left.\frac{\Delta_*}{\ADM}\right|_{\halpha=\halpha_c}
+
(\halpha-\halpha_c) \frac{d}{d\halpha} \left[ \frac{\Delta_*}{\ADM} \right]_{\halpha=\halpha_c} \nonumber\\
&+& O((\halpha-\halpha_c)^2)\,,
\label{eq:DeltaMtotSeries_exp}
\eea
where the coefficient of the linear term in $\halpha-\halpha_c$ is strictly positive. 
Thus, our results from Section~\ref{subsubsec:exponential} predict that $\Delta_*/\ADM$ increases as a function of $\alpha/\ADM^2$, at least when the small hairy BH is perturbatively close to Schwarzschild. This is not apparent in the outcome of the numerical simulations of~\cite{Capuano:2026lhs}.

As a final remark, we note that if all quantities were expressed in units of the small black hole's mass, $M$, instead of the total initial mass, $\ADM$, the characterization of the merger would change significantly. For instance, the dimensionless merger duration $\Delta_*/M$ as a function of the dimensionless coupling constant $\alpha/M^2$ is always shorter than in GR. This is a consequence of $M/r_h$ departing faster from its corresponding GR value than $\Delta_*/r_h$. We have confirmed numerically this to happen with all the coupling functions considered: linear, quadratic and exponential.

\section{Conclusions and Discussion\label{sec:conclusion}}

In this paper we studied the merger phase of two non-spinning hairy black holes, in the context of EsGB gravity and taking the extreme mass ratio limit.
Using ray-tracing techniques valid in this regime, we characterized the evolution of the event horizon during a head-on plunge. In particular, we quantified the merger duration and the area increase of the small BH as a function of the coupling constant of the theory.

Our results indicate that, for the three EsGB models studied, the plunge phase lasts longer than in GR, for sufficiently small coupling constants $\alpha$ that still allow for hairy black holes. However, for the model featuring a special exponential coupling, the merger duration reveals a non-monotonic behavior, and for $\alpha/r_h^2 \gsim 7$ it becomes shorter than in GR. This feature makes this model particularly interesting for the analysis of the whole merger since it suggests that both longer or shorter merging times are possible, depending on the coupling constant. These results hold when normalizing all quantities with respect to the horizon radius of the small BH, $r_h$.
We speculate that numerical simulations of this model, as those of Ref.~\cite{Capuano:2026lhs}, might also identify a non-monotonic behavior in the duration of the merger phase.

The longer plunging times we found for EsGB mergers (for sufficiently small $\alpha$) are not generic among modified theories of gravity. In fact, by using the same techniques, the merger duration in Einsteinian cubic gravity~\cite{Bueno:2016xff} (which is a particular example of a higher derivative theory that has a well-posed initial value formulation~\cite{Figueras:2024bba}) was found to be shorter than in GR, for all values of the coupling constant studied, and again keeping $r_h$ fixed~\cite{Dias:2024wib}.

We have now accumulated considerable evidence, both in this paper and in \cite{Dias:2024wib}, that the merger phase duration and the area increment both track the behavior of the photon ring of the small black hole, as a function of the coupling constant. 
It is well known that the properties of the photon ring control the QNM ringdown~\cite{Ferrari:1984zz,Mashhoon:1985cya,Cardoso:2008bp}. Our findings indicate that the photon ring also has important bearing on the merger dynamics. 
This aligns well with recent work arguing that the observed simplicity and universality of merger waveforms may be governed by strong-field mechanisms, thus motivating analytic control~\cite{Jaramillo:2023day}.
It would be interesting to know if similar behavior is obtained for other black hole horizons, such as the apparent horizon and other marginally outer trapped surfaces considered in the literature~\cite{Pook-Kolb:2019ssg, Pook-Kolb:2020zhm, Pook-Kolb:2020jlr, Booth:2020qhb, Pook-Kolb:2021gsh}, for these modified theories of gravity.

To compare with NR simulations we converted our results (normalized with respect to $r_h$) to express all quantities in units of the total initial mass of the binary, $\ADM$. Within the regime of validity of our calculations, we find that normalizing with respect to $r_h$ is a good indicator of the behavior of the merger duration as a function of the coupling constant if we had instead normalized with respect to $\ADM$. The inclusion of estimates of the radiated energy to infinity yields negligible corrections in the extreme mass ratio regime.

Black hole mergers are among the best physical systems to study and constrain modified theories of gravity, such as the EsGB models. While perturbative approaches have been applied in this context for more than a decade to address the inspiral and ringdown phases, the first firm steps are being now taken to evolve these systems numerically through the non-linear merger phase. The methods we have employed complement those of numerical simulations, while allowing for much faster exploration of the parameter space of modified theories. We expect that a two-way exchange of ideas between NR studies and alternative approaches such as this one will play a key role in advancing the field.

\begin{acknowledgments}

We thank Enrico Barausse and Masato Minamitsuji for helpful discussions and correspondence.
AMF and JVR are thankful to the Institute for Fundamental Physics of the Universe (IFPU) and the Gravity Theory Trust (GTT) for hospitality and financial support of the focus week program “The Dynamics of Black Hole Mergers and Gravitational Wave Generation”, where part of this work was completed.
AMF acknowledges the support from EU Horizon 2020 Research and Innovation Programme under the Marie Sk\l{}odowska-Curie Grant Agreement no. 101149470.
DCL and JVR acknowledge support from  {\it Funda\c{c}\~ao para a Ci\^encia e a Tecnologia} -- FCT, Portugal, under project No. 2022.08368.PTDC and for the financial support to the Center for Astrophysics and Gravitation (CENTRA/IST/ULisboa)
through grants UID/PRR/00099/2025 and UID/
00099/2025. 
DCL thanks the University of Hertfordshire for a PhD studentship.
JVR acknowledges the support from {\it Funda\c{c}\~ao para a Ci\^encia e a Tecnologia} -- FCT, Portugal, under project No. 2024.04456.CERN. 
\end{acknowledgments}

\appendix
\section{EsGB black hole solutions \label{app:f_solutions}}

\begin{figure*}[t!]
    \centering
    \begin{minipage}{0.33\linewidth}
        \centering
        \includegraphics[width=\linewidth]{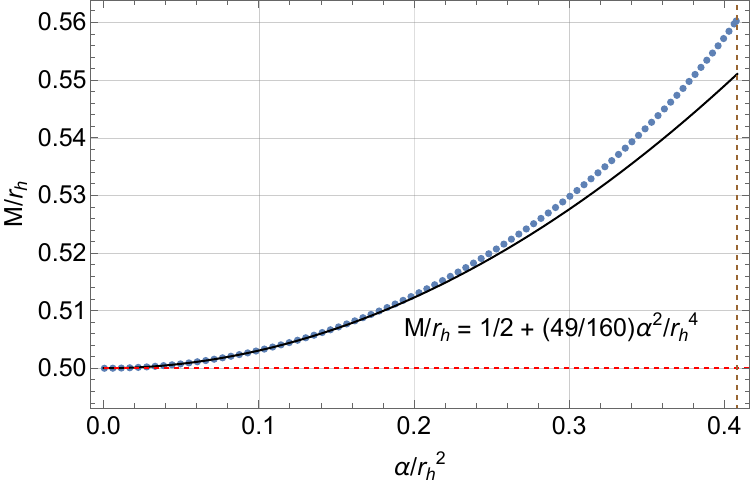}
    \end{minipage}\hfill
    \begin{minipage}{0.33\linewidth}
        \centering
        \includegraphics[width=\linewidth]{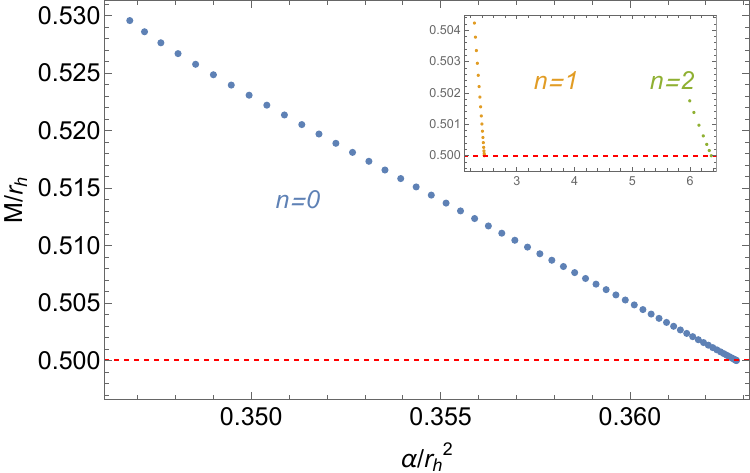}
    \end{minipage}
    \begin{minipage}{0.33\linewidth}
        \centering
        \includegraphics[width=\linewidth]{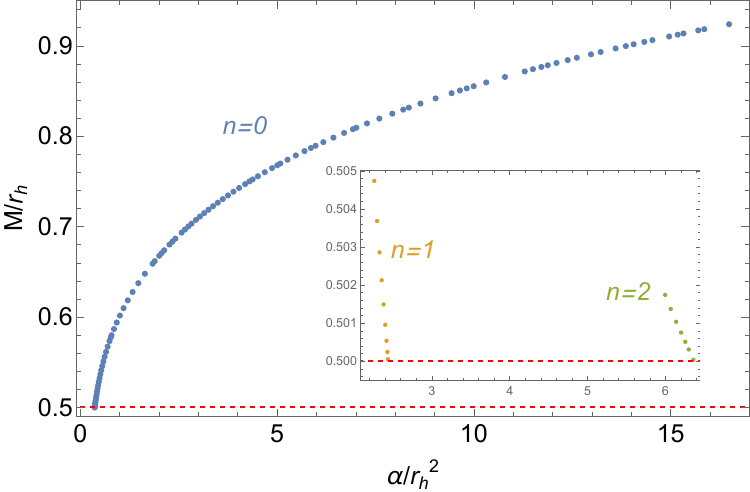}
    \end{minipage}
    \caption{The ratio between mass $M$ and horizon radius $r_h$ as a function of the dimensionless coupling constant $\alpha/r_h^2$ for the EsGB BHs with the different coupling functions studied: linear, shown only for $\alpha\geq0$ (left), quadratic (middle) and exponential (right). The middle and right panels show the fundamental branch ($n=0$) of EsGB BHs solution, while branches with $n=1$ and $n=2$ nodes in their radial profile are displayed in the inset.}
    \label{fig:mass_vs_alpha}
\end{figure*}

In this Appendix we provide a few details about the numerical construction of the static hairy BHs described in Section~\ref{sec:EsGB}.
In the shift-symmetric theory these spacetimes necessarily depart from the vacuum solutions of GR~\cite{Sotiriou:2014pfa}. In other cases, in particular with the quadratic or the special exponential coupling, the Schwarzschild geometry is also a solution, but it suffers from a tachyonic instability induced when the Kretschmann scalar exceeds a critical value, thus triggering the appearance of BH solutions with non-trivial scalar hair~\cite{Doneva:2017bvd,Silva:2017uqg}.
All of these solutions can be obtained by numerically integrating Eqs.~\eqref{eq:MFE1}-\eqref{eq:SFE}.

Start with Eq.~\eqref{eq:MFE2}, which can be readily solved for $e^{\Lambda}$, giving
\be
\label{eq:lambda}
    e^{\Lambda}= \frac{-A\pm\sqrt{A^2-4B}}{2}\,,
\ee
with $A= -1+\frac{1}{4}r^2\Phi'^2-\Gamma'(r+2\alpha\dot{f}\Phi')$, $B = 6\alpha\dot{f}\Gamma'\Phi'$.
In Eq.~\eqref{eq:lambda}, the plus sign must be chosen to ensure that $e^{-\Lambda(r)}|_{r= r_h} = 0$, as it should for a BH solution \cite{Antoniou:2017acq,Antoniou_2018_1}. In turn, regularity of $\Phi''(r_h)$ provides a relation between $\Phi'(r_h)$ and $\Phi(r_h)$,
\be
\label{eq:derphi_cond}
    \Phi'(r_h)=\frac{r_h}{4\alpha\dot{f}(\Phi(r_h))}\left(-1\pm\sqrt{1-\frac{96 \alpha^2 \dot{f}(\Phi(r_h))^2}{r_h^4}}\right)\,.
\ee
We must choose the plus sign in Eq.~\eqref{eq:derphi_cond} to guarantee that the Schwarzschild solution can be obtained from the $\Phi(r_h)\rightarrow 0$ limit. Additionally, solutions must obey the condition 
\be
\label{eq:inequality}
    \frac{96\alpha^2\dot{f}(\Phi(r_h))^2}{r_h^4} \leq 1
\ee
to guarantee reality of~\eqref{eq:derphi_cond}.

Using Eq.~\eqref{eq:lambda}, the system of Eqs.~\eqref{eq:MFE1},~\eqref{eq:MFE3} and ~\eqref{eq:SFE} reduces to a coupled system of two second-order differential equations for $\Phi$ and $\Gamma$. In practice, none of the equations depends on $\Gamma$ explicitly, only on its derivatives. Thus, we can solve the system for $\Phi$ and $\Gamma'$ and integrate the latter independently to obtain $\Gamma$. Near the horizon we have $e^{\Gamma(r)}\sim (r-r_h) + \mathcal{O}(r-r_h)^2$, so that $\Gamma'(r) = 1/(r-r_h)+\mathcal{O}(1)$. For numerical convenience, we define the function $\Psi(r)=\Gamma'(r)(r-r_h)$ and integrate the system of equations for the functions $\Phi$ and $\Psi$ up to a large radius, $r_f$, giving as initial conditions~\cite{Kokkotas_2017}
\bea
    &\Phi(r_i) = \Phi_h\, \,, \, \Phi'(r_i)= \Phi'_h \, , \,    &\Psi(r_i) = 1\,,
\eea
where $r_i = r_h(1+\epsilon)$, $\epsilon \ll 1$, $\Phi_h$ is a free parameter, and $\Phi'_h$ is given by Eq.~\eqref{eq:derphi_cond} with the appropriate substitution by $r_h \to r_i$. Close to $r_f$, the metric functions and the scalar field are given by their asymptotic behavior, i.e. $e^{\Gamma (r)} = 1-2M/r+\mathcal{O}(1/r^2)$, $e^{\Lambda (r)} = 1+2M/r+\mathcal{O}(1/r^2)$, and $\Phi (r) = \Phi_\infty+ D/r + \mathcal{O}(1/r^2)$, where $M$, $\Phi_\infty$ and $D$ are, respectively, the mass of the black hole, the value of the scalar field at infinity, and the charge of the scalar field. 
In turn, the function $\Psi$ behaves asymptotically as $\Psi(r) = 2M/r+\mathcal{O}(1/r^2)$. The values for $M$, $\Phi_\infty$ and $D$ can be obtained by matching the asymptotic behavior of $\Psi$ and $\Phi$ to their numerical solutions close to $r_f$. Note that these quantities will generically depend on the choice of $\alpha$.
Once $\Psi$ is computed, $e^\Gamma$ can be obtained through
\be
    e^{\Gamma(r)}=\left(1-\frac{2M}{r_f}\right)\exp\left(-\int_{r}^{r_f} \frac{\Psi(r')}{r'-r_h} dr'\right)\,,
\ee
while $\Phi$ and $\Gamma$ determine $e^\Lambda$ according to Eq.~\eqref{eq:lambda}.

We demand an asymptotically flat scalar field, with $\Phi_\infty = 0$. This last condition selects specific values of $(\Phi_h$, $\alpha/r_h^2)$, for which asymptotically flat black hole solutions endowed with non-trivial scalar hair exist. 
In the case of a linear coupling function, there is only a 1-parameter family of regular BH solutions, as shown in~\cite{Sotiriou:2014pfa}. Once the horizon radius (or equivalently, the mass $M$) is fixed, regularity allows only one value of $\Phi_h$ (or correspondingly, the scalar charge $D$) for any given $\alpha$. Therefore, hairy BHs in shift-symmetric EsGB gravity fall along a single line in the $(\Phi_h$, $\alpha/r_h^2)$ plane.

When adopting the quadratic or the special exponential coupling function, the space of solutions becomes richer. The Schwarzschild geometry $\Phi_h = 0$ is still a solution of the EsGB field equations in those cases, independently of the value of $\alpha/r_h^2$. Nevertheless, it becomes unstable for specific discrete values of $\alpha/r_h^2$, triggering BH solutions with $\Phi_h \neq 0$ (and thus non-trivial scalar hair since $\Phi_\infty=0$). These `spontaneously scalarized' solutions exist in branches, starting at $\Phi_h = 0$ and some critical $\alpha/r_h^2$, and developing in the $(\Phi_h,\alpha/r_h^2)$-plane until the inequality in Eq.~\eqref{eq:inequality} is no longer satisfied. The different countable solution branches correspond to distinct scalar field profiles --- solutions along the $n=0$ branch exhibit a scalar field profile with zero nodes, solutions on the $n=1$ branch have a profile with one node, and so on. While there should be an infinite number of branches, associated with the different modes triggering the tachyonic instability of the Schwarzschild solution, they increasingly resemble their GR counterpart as the range of $\Phi_h$ allowed by Eq.~\eqref{eq:inequality} narrows.

In Fig.~\ref{fig:mass_vs_alpha} we show the dependence of the mass $M$ on the coupling constant $\alpha$, both expressed in units of $r_h$, for the three EsGB models considered. These results are used in Section~\ref{sec:comparison} and in Appendices~\ref{app:details_comp} and~\ref{app:details_comp_exp}.

\section{Comparison to simulations with linear coupling\label{app:details_comp}}

This Appendix gathers some details about the calculations reported in Section~\ref{sec:comparison}, including loss of energy in the binary system due to scalar and gravitational radiation to infinity. This energy loss, $E$, enters the expression for the total mass of the final (merged) BH, $\ADM$, in the following way:
\be
\label{eq:MtotAlpha}
\ADM(\alpha) = M_\infty + M(\alpha) - E(\alpha)\,.
\ee
Recall that $M_\infty$ represents the ADM mass of the larger BH and is independent of $\alpha$. On the other hand, the ADM mass of the scalarized smaller companion, $M$, and the total energy radiated, $E$, are generically $\alpha$-dependent quantities.
If one neglects the radiated energy, this expression reduces to the one adopted in Section~\ref{sec:comparison}.

To derive equation~\eqref{eq:FinalMassModGrav}  in the main text, we begin by comparing the total mass~\eqref{eq:MtotAlpha} to the corresponding relation in the case of GR (i.e., setting $\alpha=0$), 
\be
\label{eq:MtotZero}
\ADM(0) = M_\infty + M(0) - E(0)\,.
\ee
From \eqref{eq:MtotAlpha} and \eqref{eq:MtotZero} it directly follows that
\be
\label{eq:Malpha}
\ADM(\alpha) = \ADM(0) + \delta M(\alpha) - \delta E(\alpha)\,,
\ee
where $\delta M(\alpha) = M(\alpha) - M(0)$ and $\,\delta E(\alpha) = E(\alpha) - E(0)$.

The energy emitted by extreme mass ratio BH binaries in scalar-tensor theories, $E(\alpha)$, has been studied in~\cite{Maselli:2020zgv}.
Their results indicate that the relative difference between the radiated energy flux in EsGB and in GR is proportional to the square of the dimensionless scalar charge $d$ of the small BH:
\be
   \frac{\delta \dot{E} (\alpha)}{\dot{E}(0)} = \varepsilon\, d^2\,,
   \label{eq:flux}
\ee
where the overdot denotes a time derivative, and \cite{Maselli:2020zgv, Sotiriou:2014pfa}
\be
d = \frac{2\alpha}{M(0)^2} + O(\alpha^3)\,.
\label{eq:d_def}
\ee
The proportionality factor $\varepsilon$ depends on the distance between the binary's BHs and satisfies $0<\varepsilon<1$, becoming smaller as the distance decreases (see Fig.~1 of~\cite{Maselli:2020zgv}).

Combining Eqs.~\eqref{eq:Malpha}, \eqref{eq:MassAlpha} and the integrated version of~\eqref{eq:flux}, one gets
\be
    \ADM(\talpha) = \ADM(0) 
    + \frac{49}{160} r_h \talpha^2
    - E(0) \varepsilon\, d^2
    +O(\talpha^3)\,,
    \label{eq:M_fin_alpha}
\ee
where $\talpha$ was defined in~\eqref{eq:alphatilde_def}.
Plugging Eq. \eqref{eq:d_def} into \eqref{eq:M_fin_alpha}
and dividing both sides by $r_h$, we obtain Eq.~\eqref{eq:FinalMassModGrav}, with the constants $b$ and $c$ given explicitly by
\be
    b=\frac{\ADM(0)}{r_h}\,,  
    \qquad   
    c=\frac{49}{160} - 32\,\varepsilon \frac{E(0)}{M(0)} \,.
\label{eq:bc}
\ee
In simplifying the expression for $c$ we used the relation $r_h=2M$, valid for $\alpha=0$.
Given that, in the extreme mass ratio regime, one has $E(0)/M(0) \simeq 0.01\, M(0)/M_\infty\ll 10^{-2}$ \cite{Davis:1971gg}, and that $0<\varepsilon<1$, we conclude that $c$ is positive and that the contribution from the energy radiated to this expression (linear in $\varepsilon$) is negligible compared to the other contribution at the same quadratic order in $\talpha$.

To close this Appendix, we turn to the derivation of Eq.~\eqref{eq:DeltaMtotSeries}, starting from~\eqref{eq:DeltarhSeries} and~ \eqref{eq:FinalMassModGrav}.  
In order to characterize the behavior of $\Delta_*/\ADM$ when $\talpha$ is small, we perform a Taylor expansion around $\talpha=0$:
\bea
\label{eq:DeltaMtotSeries2}
\frac{\Delta_*}{\ADM}(\talpha) &=&
\frac{\Delta_*}{\ADM}(0) 
+ \talpha \frac{d}{d\talpha} \left[ \frac{\Delta_*}{\ADM} \right]_{\talpha=0} \\
&+&  \frac{\talpha^2}{2} \frac{d^2}{d\talpha^2} \left[ \frac{\Delta_*}{\ADM} \right]_{\talpha=0} 
+ O(\talpha^3)\,.
\nonumber
\eea
Given that the first derivative of $\Delta_*/r_h$ with respect to $\talpha$ vanishes at $\talpha=0$ (see Eq.~\eqref{eq:DeltarhSeries}), we have
\begin{flalign}
&\frac{d}{d\talpha} \left[ \frac{\Delta_*}{\ADM} \right]_{\talpha=0} 
= 
\frac{d}{d\talpha} \left[ \frac{\Delta_*}{r_h} \left(\frac{\ADM}{r_h}\right)^{-1} \right]_{\talpha=0}\\
& = \frac{1}{b} \frac{d}{d\talpha} \left[ \frac{\Delta_*}{r_h} \right]_{\talpha=0} - \left.\frac{2c\, \talpha}{b^2} \frac{\Delta_*}{r_h} \right|_{\talpha=0}=0\,,\nonumber
\end{flalign}
where we used Eq.~\eqref{eq:FinalMassModGrav} in the second step.
Evaluating the second derivative, we get
\be
\frac{d^2}{d\talpha^2} \left[ \frac{\Delta_*}{\ADM} \right]_{\talpha=0} 
= 
\frac{1}{b} \frac{d^2}{d\talpha^2} \left[ \frac{\Delta_*}{r_h} \right]_{\talpha=0} 
 - \frac{2c}{b^2} \left.\frac{\Delta_*}{r_h}\right|_{\talpha=0} \,.
 \label{eq:2nd_der_linear}
\ee
Once again, in the extreme mass ratio, this expression formally vanishes. Yet, the second contribution is suppressed compared to the first term by a factor of order $O(M/M_\infty)\ll 1$. 
Therefore, the second derivative is strictly positive and we conclude that $\Delta_*/\ADM$ is locally an increasing function of $\talpha$, for small enough $\talpha$.

At this point let us discuss what would happen if we normalized $\Delta_*$ with respect to the individual mass $M$ instead of $\ADM$. The linear term in the expansion of $\Delta_*/M$, equivalent to~\eqref{eq:DeltaMtotSeries2}, still vanishes, but the quadratic term instead is negative in this case.
This can be seen by using Eqs.~\eqref{eq:DeltarhSeries} and~\eqref{eq:bc} to evaluate~\eqref{eq:2nd_der_linear}. The numerical results shown in Fig.~\ref{fig:overall_linear} can be represented using units of $M$, and we find that it agrees with this analysis.

Finally, we can convert the expansion~\eqref{eq:DeltaMtotSeries2} into an equivalent expansion in terms of the variable $\halpha$. The connection with $\talpha$ is provided by~\eqref{eq:FinalMassModGrav}, which yields
\be
\halpha = \frac{\talpha}{b^2} + O(\talpha^3)\,.
\ee
With this, we can evaluate
\be
\frac{d}{d\halpha} \left[ \frac{\Delta_*}{\ADM} \right]_{\halpha=0} 
=
b^2 \frac{d}{d\talpha} \left[ \frac{\Delta_*}{\ADM} \right]_{\talpha=0} = 0\,,
\ee
and
\be
\frac{d^2}{d\halpha^2} \left[ \frac{\Delta_*}{\ADM} \right]_{\halpha=0} 
=
b^4 \frac{d^2}{d\talpha^2} \left[ \frac{\Delta_*}{\ADM} \right]_{\talpha=0} > 0\,,
\ee
yielding Eq.~\eqref{eq:DeltaMtotSeries} of the main text.

\section{Comparison to simulations with exponential coupling\label{app:details_comp_exp}}

In this Appendix we repeat the analysis of Appendix~\ref{app:details_comp} for the case of the special exponential coupling case \eqref{eq:exp_coupling}, where $f(\Phi)=\frac{1}{6}(1-e^{-3 \Phi^2 /2})\,$. The main difference is that all series expansions will be performed around $\talpha=\talpha_c=0.362811$ (where the scalarized BHs with $n=0$ branch off from Schwarzschild) instead of $\talpha=0$ and only up to first order.

Equation \eqref{eq:Malpha} is still valid if one shifts $\talpha=0$ to $\talpha=\talpha_c$:
\be
    \ADM(\talpha) = \ADM(\talpha_c) + \delta M(\talpha) - \delta E(\talpha)\,,
    \label{eq:MfinExp}
\ee
where now $\delta M(\talpha) = M(\talpha) - M(\talpha_c)$ and $\delta E(\talpha) = E(\talpha) - E(\talpha_c) = E(\talpha_c) \varepsilon d^2$.
As before, we need the expressions to replace $\delta M (\talpha)$ and $d^2$. We will see below that both of these are linear in $(\talpha-\talpha_c)$.

By fitting the results displayed in the third panel of Fig.~\ref{fig:mass_vs_alpha}, we obtain
\be
 M(\talpha) = M(\talpha_c) + 0.3\, r_h (\talpha-\talpha_c) + O((\talpha-\talpha_c)^2)\,.
 \label{eq:MExp}
\ee
Next we turn to $d=D/M$, where $D$ is the dimensionful scalar charge of the small BH (see Appendix~\ref{app:f_solutions}). 
From fitting our numerical results in~\ref{subsubsec:exponential} we find that:
\be
d= \frac{D}{M} = 1.65\sqrt{\talpha-\talpha_c} + O((\talpha-\talpha_c)^{3/2})\,,
\label{eq:dtalpha}
\ee
which is in very good agreement with Fig. 4 of \cite{Doneva:2017bvd} once the appropriate conversions are made.
Therefore, close to the critical coupling $\talpha_c$, $d^2$ grows linearly in $(\talpha-\talpha_c)$, as previously advocated.

Combining \eqref{eq:MfinExp}, \eqref{eq:MExp} and \eqref{eq:dtalpha}, we get Eq.~\eqref{eq:FinalMassModGrav_exp} in the main text, with
\be
    \beta = \frac{\ADM(\talpha_c)}{r_h}\,,
    \qquad 
    \gamma = 0.3 - (1.65)^2 \frac{E(\talpha_c)}{2 M(\talpha_c)} \varepsilon\,.
    \label{eq:value_gamma}
\ee
As in Appendix~\ref{app:details_comp}, the second term in the expression of $\gamma$ is suppressed compared to the first one, in the EMR limit. 
In this case, $\ADM/r_h$ grows linearly as a function of $\talpha$ when $\talpha\simeq\talpha_c$.

In order to characterize the behavior of $\Delta_*/\ADM$ of spontaneously scalarized BHs that are perturbatively close to Schwarzschild, we perform a Taylor expansion around $\talpha\simeq\talpha_c$:
\bea
\frac{\Delta_*}{\ADM}(\talpha) &=&
\frac{\Delta_*}{\ADM}(\talpha_c) 
+ (\talpha-\talpha_c) \frac{d}{d\talpha} \left[ \frac{\Delta_*}{\ADM} \right]_{\talpha=\talpha_c}\nonumber \\
&+&  O((\talpha-\talpha_c)^2)\,.
\label{eq:DeltaMtotSeries_exp2}
\eea
From Eq.~\eqref{eq:DeltarhSeries_exp} and Eq.~\eqref{eq:FinalMassModGrav_exp} we can compute the first derivative:
\begin{flalign}
&\frac{d}{d\talpha} \left[ \frac{\Delta_*}{\ADM} \right]_{\talpha=\talpha_c} 
= 
\frac{d}{d\talpha} \left[ \frac{\Delta_*}{r_h} \left(\frac{\ADM}{r_h}\right)^{-1} \right]_{\talpha=\talpha_c}\nonumber\\
& = \frac{1}{\beta} \frac{d}{d\talpha} \left[ \frac{\Delta_*}{r_h} \right]_{\talpha=\talpha_c} - \left.\frac{\gamma}{\beta^2} \frac{\Delta_*}{r_h} \, \right|_{\talpha=\talpha_c}\,.
\end{flalign}

Once again, in the extreme mass ratio, this expression formally vanishes. Yet, the second contribution is suppressed compared to the first term by a factor of order $O(M/M_\infty)\ll 1$. 
Therefore, we conclude that $\Delta_*/\ADM$ is locally an increasing function of $\talpha$, for $\talpha\simeq\talpha_c$.

Finally, we convert the expansion~\eqref{eq:DeltaMtotSeries_exp2} into an equivalent expansion in terms of the variable $\halpha$. In this case, the connection with $\talpha$ is provided by~\eqref{eq:FinalMassModGrav_exp}, which yields
\be
\halpha = \frac{\talpha}{\beta^2} + O(\talpha-\talpha_c)\,.
\ee
Thus, we obtain
\be
\frac{d}{d\halpha} \left[ \frac{\Delta_*}{\ADM} \right]_{\halpha=\halpha_c} 
=
\beta^2 \frac{d}{d\talpha} \left[ \frac{\Delta_*}{\ADM} \right]_{\talpha=\talpha_c} >0\,,
\ee
yielding Eq.~\eqref{eq:DeltaMtotSeries_exp} of the main text.

\bibliography{BHmergers_in_EsGB_refs}

@misc{LIGOScientific:2025snk,
    author = "Abac, A. G. and others",
    collaboration = "LIGO Scientific, VIRGO, KAGRA",
    title = "{Open Data from LIGO, Virgo, and KAGRA through the First Part of the Fourth Observing Run}",
    eprint = "2508.18079",
    archivePrefix = "arXiv",
    primaryClass = "gr-qc",
    reportNumber = "LIGO-P2500167",
    month = "8",
    year = "2025"
}

@misc{LIGOScientific:2025bkz,
    author = "Abac, A. G. and others",
    collaboration = "LIGO Scientific, VIRGO, KAGRA",
    title = "{Directional Search for Persistent Gravitational Waves: Results from the First Part of LIGO-Virgo-KAGRA's Fourth Observing Run}",
    eprint = "2510.17487",
    archivePrefix = "arXiv",
    primaryClass = "gr-qc",
    reportNumber = "LIGO-P250038",
    month = "10",
    year = "2025"
}

@ARTICLE{2017arXiv170200786A,
       author = {{Amaro-Seoane}, P. and others},
        title = "{Laser Interferometer Space Antenna}",
      journal = {arXiv e-prints},
         year = 2017,
        month = feb,
          eid = {arXiv:1702.00786},
        pages = {arXiv:1702.00786},
          doi = {10.48550/arXiv.1702.00786},
archivePrefix = {arXiv},
       eprint = {1702.00786},
 primaryClass = {astro-ph.IM},
       adsurl = {https://ui.adsabs.harvard.edu/abs/2017arXiv170200786A}
}

@article{Punturo_2010,
doi = {10.1088/0264-9381/27/19/194002},
url = {https://doi.org/10.1088/0264-9381/27/19/194002},
year = {2010},
month = {sep},
publisher = {},
volume = {27},
number = {19},
pages = {194002},
author = {Punturo, M. and others},
title = {The Einstein Telescope: a third-generation gravitational wave observatory},
journal = {Classical and Quantum Gravity},
}

@article{ET:2019dnz,
    author = "Maggiore, Michele and others",
    collaboration = "ET",
    title = "{Science Case for the Einstein Telescope}",
    eprint = "1912.02622",
    archivePrefix = "arXiv",
    primaryClass = "astro-ph.CO",
    doi = "10.1088/1475-7516/2020/03/050",
    journal = "JCAP",
    volume = "03",
    pages = "050",
    year = "2020"
}

@article{Reitze:2019iox,
    author = "Reitze, David and others",
    title = "{Cosmic Explorer: The U.S. Contribution to Gravitational-Wave Astronomy beyond LIGO}",
    eprint = "1907.04833",
    archivePrefix = "arXiv",
    primaryClass = "astro-ph.IM",
    reportNumber = "LIGO-P1900316",
    journal = "Bull. Am. Astron. Soc.",
    volume = "51",
    number = "7",
    pages = "035",
    year = "2019"
}

@article{LISA:2022kgy,
    author = "Arun, K. G. and others",
    collaboration = "LISA",
    title = "{New horizons for fundamental physics with LISA}",
    eprint = "2205.01597",
    archivePrefix = "arXiv",
    primaryClass = "gr-qc",
    doi = "10.1007/s41114-022-00036-9",
    journal = "Living Rev. Rel.",
    volume = "25",
    number = "1",
    pages = "4",
    year = "2022"
}

@article{Barack:2018yly,
    author = "Barack, Leor and others",
    title = "{Black holes, gravitational waves and fundamental physics: a roadmap}",
    eprint = "1806.05195",
    archivePrefix = "arXiv",
    primaryClass = "gr-qc",
    doi = "10.1088/1361-6382/ab0587",
    journal = "Class. Quant. Grav.",
    volume = "36",
    number = "14",
    pages = "143001",
    year = "2019"
}

@article{Gnocchi:2019jzp,
    author = "Gnocchi, G. and others",
    title = "{Bounding alternative theories of gravity with multiband GW observations}",
    eprint = "1905.13460",
    archivePrefix = "arXiv",
    primaryClass = "gr-qc",
    doi = "10.1103/PhysRevD.100.064024",
    journal = "Phys. Rev. D",
    volume = "100",
    number = "6",
    pages = "064024",
    year = "2019"
}

@article{Baker:2014zba,
    author = "Baker, Tessa and Psaltis, Dimitrios and Skordis, Constantinos",
    title = "{Linking Tests of Gravity On All Scales: from the Strong-Field Regime to Cosmology}",
    eprint = "1412.3455",
    archivePrefix = "arXiv",
    primaryClass = "astro-ph.CO",
    doi = "10.1088/0004-637X/802/1/63",
    journal = "Astrophys. J.",
    volume = "802",
    pages = "63",
    year = "2015"
}

@misc{LIGOScientific:2025obp,
    author = "{LIGO Scientific Collaboration, Virgo Collaboration, KAGRA Collaboration}",
    title = "{Black Hole Spectroscopy and Tests of General Relativity with GW250114}",
    eprint = "2509.08099",
    archivePrefix = "arXiv",
    primaryClass = "gr-qc",
    reportNumber = "LIGO P2500461",
    month = "9",
    year = "2025"
}

@article{LIGOScientific:2025rid,
    author = "Abac, A. G. and others",
    collaboration = "LIGO Scientific, Virgo, KAGRA",
    title = "{GW250114: Testing Hawking{\textquoteright}s Area Law and the Kerr Nature of Black Holes}",
    eprint = "2509.08054",
    archivePrefix = "arXiv",
    primaryClass = "gr-qc",
    reportNumber = "LIGO-P2500421",
    doi = "10.1103/kw5g-d732",
    journal = "Phys. Rev. Lett.",
    volume = "135",
    number = "11",
    pages = "111403",
    year = "2025"
}

@misc{Corman:2025wun,
    author = "Corman, Maxence and Arest{\'e} Sal{\'o}, Llibert and Clough, Katy",
    title = "{Black hole binaries in shift-symmetric Einstein-scalar-Gauss-Bonnet gravity experience a slower merger phase}",
    eprint = "2511.19073",
    archivePrefix = "arXiv",
    primaryClass = "gr-qc",
    month = "11",
    year = "2025"
}

@article{Elder:2022rak,
    author = "Elder, Benjamin and Sakstein, Jeremy",
    title = "{Mapping the weak field limit of scalar-Gauss-Bonnet gravity}",
    eprint = "2210.10955",
    archivePrefix = "arXiv",
    primaryClass = "gr-qc",
    doi = "10.1103/PhysRevD.107.044006",
    journal = "Phys. Rev. D",
    volume = "107",
    number = "4",
    pages = "044006",
    year = "2023"
}

@article{Amendola:2007ni,
    author = "Amendola, Luca and Charmousis, Christos and Davis, Stephen C.",
    title = "{Solar System Constraints on Gauss-Bonnet Mediated Dark Energy}",
    eprint = "0704.0175",
    archivePrefix = "arXiv",
    primaryClass = "astro-ph",
    doi = "10.1088/1475-7516/2007/10/004",
    journal = "JCAP",
    volume = "10",
    pages = "004",
    year = "2007"
}

@article{Yagi:2011xp,
    author = "Yagi, Kent and Stein, Leo C. and Yunes, Nicol{\'a}s and Tanaka, Takahiro",
    title = "{Post-Newtonian, Quasi-Circular Binary Inspirals in Quadratic Modified Gravity}",
    eprint = "1110.5950",
    archivePrefix = "arXiv",
    primaryClass = "gr-qc",
    doi = "10.1103/PhysRevD.85.064022",
    journal = "Phys. Rev. D",
    volume = "85",
    pages = "064022",
    year = "2012",
    note = "[Erratum: Phys.Rev.D 93, 029902 (2016)]"
}

@article{Kanti:1995vq,
    author = "Kanti, P. and Mavromatos, N. E. and Rizos, J. and Tamvakis, K. and Winstanley, E.",
    title = "{Dilatonic black holes in higher curvature string gravity}",
    eprint = "hep-th/9511071",
    archivePrefix = "arXiv",
    reportNumber = "CERN-TH-95-297, OUTP-95-43-P, CERN-TH/95-297, OUTP-95-43P",
    doi = "10.1103/PhysRevD.54.5049",
    journal = "Phys. Rev. D",
    volume = "54",
    pages = "5049--5058",
    year = "1996"
}

@article{Chung:2024vaf,
    author = "Chung, Adrian Ka-Wai and Yunes, Nicolas",
    title = "{Quasinormal mode frequencies and gravitational perturbations of black holes with any subextremal spin in modified gravity through METRICS: The scalar-Gauss-Bonnet gravity case}",
    eprint = "2406.11986",
    archivePrefix = "arXiv",
    primaryClass = "gr-qc",
    doi = "10.1103/PhysRevD.110.064019",
    journal = "Phys. Rev. D",
    volume = "110",
    number = "6",
    pages = "064019",
    year = "2024"
}

@article{Khoo:2024agm,
    author = "Khoo, Fech Scen and Bl{\'a}zquez-Salcedo, Jose Luis and Kleihaus, Burkhard and Kunz, Jutta",
    title = "{Quasinormal modes of rotating black holes in shift-symmetric Einstein-scalar-Gauss{\textendash}Bonnet theory}",
    eprint = "2412.09377",
    archivePrefix = "arXiv",
    primaryClass = "gr-qc",
    doi = "10.1140/epjc/s10052-025-15106-9",
    journal = "Eur. Phys. J. C",
    volume = "85",
    number = "11",
    pages = "1366",
    year = "2025"
}

@article{Sotiriou:2013qea,
    author = "Sotiriou, Thomas P. and Zhou, Shuang-Yong",
    title = "{Black hole hair in generalized scalar-tensor gravity}",
    eprint = "1312.3622",
    archivePrefix = "arXiv",
    primaryClass = "gr-qc",
    doi = "10.1103/PhysRevLett.112.251102",
    journal = "Phys. Rev. Lett.",
    volume = "112",
    pages = "251102",
    year = "2014"
}

@article{Sotiriou:2014pfa,
    author = "Sotiriou, Thomas P. and Zhou, Shuang-Yong",
    title = "{Black hole hair in generalized scalar-tensor gravity: An explicit example}",
    eprint = "1408.1698",
    archivePrefix = "arXiv",
    primaryClass = "gr-qc",
    doi = "10.1103/PhysRevD.90.124063",
    journal = "Phys. Rev. D",
    volume = "90",
    pages = "124063",
    year = "2014"
}

@article{Delgado:2020rev,
    author = "Delgado, Jorge F. M. and Herdeiro, Carlos A. R. and Radu, Eugen",
    title = "{Spinning black holes in shift-symmetric Horndeski theory}",
    eprint = "2002.05012",
    archivePrefix = "arXiv",
    primaryClass = "gr-qc",
    doi = "10.1007/JHEP04(2020)180",
    journal = "JHEP",
    volume = "04",
    pages = "180",
    year = "2020"
}

@article{Lyu:2022gdr,
    author = "Lyu, Zhenwei and Jiang, Nan and Yagi, Kent",
    title = "{Constraints on Einstein-dilation-Gauss-Bonnet gravity from black hole-neutron star gravitational wave events}",
    eprint = "2201.02543",
    archivePrefix = "arXiv",
    primaryClass = "gr-qc",
    reportNumber = "LIGO-P2100466",
    doi = "10.1103/PhysRevD.105.064001",
    journal = "Phys. Rev. D",
    volume = "105",
    number = "6",
    pages = "064001",
    year = "2022",
    note = "[Erratum: Phys.Rev.D 106, 069901 (2022), Erratum: Phys.Rev.D 106, 069901 (2022)]"
}

@article{Julie:2024fwy,
    author = "Juli{\'e}, F{\'e}lix-Louis and Pompili, Lorenzo and Buonanno, Alessandra",
    title = "{Inspiral-merger-ringdown waveforms in Einstein-scalar-Gauss-Bonnet gravity within the effective-one-body formalism}",
    eprint = "2406.13654",
    archivePrefix = "arXiv",
    primaryClass = "gr-qc",
    doi = "10.1103/PhysRevD.111.024016",
    journal = "Phys. Rev. D",
    volume = "111",
    number = "2",
    pages = "024016",
    year = "2025"
}

@misc{Sanger:2024axs,
    author = {S{\"a}nger, Elise M. and others},
    title = "{Tests of General Relativity with GW230529: a neutron star merging with a lower mass-gap compact object}",
    eprint = "2406.03568",
    archivePrefix = "arXiv",
    primaryClass = "gr-qc",
    reportNumber = "LIGO-P2400200",
    month = "6",
    year = "2024"
}

@article{Kovacs:2020pns,
    author = "Kov{\'a}cs, {\'A}ron D. and Reall, Harvey S.",
    title = "{Well-Posed Formulation of Scalar-Tensor Effective Field Theory}",
    eprint = "2003.04327",
    archivePrefix = "arXiv",
    primaryClass = "gr-qc",
    doi = "10.1103/PhysRevLett.124.221101",
    journal = "Phys. Rev. Lett.",
    volume = "124",
    number = "22",
    pages = "221101",
    year = "2020"
}

@misc{Figueras:2024bba,
    author = "Figueras, Pau and Held, Aaron and Kov{\'a}cs, {\'A}ron D.",
    title = "{Well-posed initial value formulation of general effective field theories of gravity}",
    eprint = "2407.08775",
    archivePrefix = "arXiv",
    primaryClass = "gr-qc",
    month = "7",
    year = "2024"
}

@article{Donoghue:1994dn,
    author = "Donoghue, John F.",
    title = "{General relativity as an effective field theory: The leading quantum corrections}",
    eprint = "gr-qc/9405057",
    archivePrefix = "arXiv",
    reportNumber = "UMHEP-408",
    doi = "10.1103/PhysRevD.50.3874",
    journal = "Phys. Rev. D",
    volume = "50",
    pages = "3874--3888",
    year = "1994"
}

@article{Bueno:2016xff,
      author         = "Bueno, Pablo and Cano, Pablo A.",
      title          = "{Einsteinian cubic gravity}",
      journal        = "Phys. Rev.",
      volume         = "D94",
      year           = "2016",
      number         = "10",
      pages          = "104005",
      doi            = "10.1103/PhysRevD.94.104005",
      eprint         = "1607.06463",
      archivePrefix  = "arXiv",
      primaryClass   = "hep-th",
      SLACcitation   = "%%CITATION = ARXIV:1607.06463;%%"
}

@article{Emparan:2016ylg,
    author = "Emparan, Roberto and Mart\'inez, Marina",
    title = "{Exact Event Horizon of a Black Hole Merger}",
    eprint = "1603.00712",
    archivePrefix = "arXiv",
    primaryClass = "gr-qc",
    doi = "10.1088/0264-9381/33/15/155003",
    journal = "Class. Quant. Grav.",
    volume = "33",
    number = "15",
    pages = "155003",
    year = "2016"
}

@article{Emparan:2016ipc,
    author = "Emparan, Roberto and Mart\'inez, Marina",
    title = "{Black hole fusion made easy}",
    doi = "10.1142/S0218271816440156",
    journal = "Int. J. Mod. Phys. D",
    volume = "25",
    number = "12",
    pages = "1644015",
    year = "2016"
}

@article{Emparan:2017vyp,
    author = "Emparan, Roberto and Mart\'inez, Marina and Zilh\~ao, Miguel",
    title = "{Black hole fusion in the extreme mass ratio limit}",
    eprint = "1708.08868",
    archivePrefix = "arXiv",
    primaryClass = "gr-qc",
    doi = "10.1103/PhysRevD.97.044004",
    journal = "Phys. Rev. D",
    volume = "97",
    number = "4",
    pages = "044004",
    year = "2018"
}

@article{Emparan:2020uvt,
    author = "Emparan, Roberto and Mar\'in, Daniel",
    title = "{Precursory collapse in Neutron Star-Black Hole mergers}",
    eprint = "2004.08143",
    archivePrefix = "arXiv",
    primaryClass = "gr-qc",
    doi = "10.1103/PhysRevD.102.024009",
    journal = "Phys. Rev. D",
    volume = "102",
    number = "2",
    pages = "024009",
    year = "2020"
}

@article{Pina:2022dye,
    author = "Pina, D. Mar\'\i{}n and Orselli, M. and Pica, D.",
    title = "{Event horizon of a charged black hole binary merger}",
    eprint = "2204.08841",
    archivePrefix = "arXiv",
    primaryClass = "gr-qc",
    doi = "10.1103/PhysRevD.106.084012",
    journal = "Phys. Rev. D",
    volume = "106",
    number = "8",
    pages = "084012",
    year = "2022"
}

@article{Dias:2023pdx,
    author = "Dias, Joao M. and Frassino, Antonia M. and Paccoia, Valentin D. and Rocha, Jorge V.",
    title = "{Black hole-wormhole collisions and the emergence of islands}",
    eprint = "2304.06098",
    archivePrefix = "arXiv",
    primaryClass = "gr-qc",
    doi = "10.1103/PhysRevD.107.124056",
    journal = "Phys. Rev. D",
    volume = "107",
    number = "12",
    pages = "124056",
    year = "2023"
}

@article{Dias:2024wib,
    author = "Dias, Jo{\~a}o M. and Frassino, Antonia M. and Lopes, David C. and Paccoia, Valentin D. and Rocha, Jorge V.",
    title = "{Impact of higher derivative corrections to general relativity on black hole mergers}",
    eprint = "2407.12947",
    archivePrefix = "arXiv",
    primaryClass = "gr-qc",
    doi = "10.1103/PhysRevD.110.124025",
    journal = "Phys. Rev. D",
    volume = "110",
    number = "12",
    pages = "124025",
    year = "2024"
}

@article{Gadioux:2024jlx,
    author = "Gadioux, Maxime and Wang, Hangzhi",
    title = "{The merger of a black hole with a cosmological horizon}",
    eprint = "2412.04551",
    archivePrefix = "arXiv",
    primaryClass = "gr-qc",
    doi = "10.1088/1361-6382/adce51",
    journal = "Class. Quant. Grav.",
    volume = "42",
    number = "9",
    pages = "095010",
    year = "2025"
}

@article{Gadioux:2024tlm,
    author = "Gadioux, Maxime and Hennigar, Robie A. and Reall, Harvey S.",
    title = "{Evolution of creases on the event horizon of a black hole merger}",
    eprint = "2407.07962",
    archivePrefix = "arXiv",
    primaryClass = "gr-qc",
    doi = "10.1103/PhysRevD.110.084029",
    journal = "Phys. Rev. D",
    volume = "110",
    number = "8",
    pages = "084029",
    year = "2024"
}

@article{Gadioux:2023pmw,
    author = "Gadioux, Maxime and Reall, Harvey S.",
    title = "{Creases, corners, and caustics: Properties of nonsmooth structures on black hole horizons}",
    eprint = "2303.15512",
    archivePrefix = "arXiv",
    primaryClass = "gr-qc",
    doi = "10.1103/PhysRevD.108.084021",
    journal = "Phys. Rev. D",
    volume = "108",
    number = "8",
    pages = "084021",
    year = "2023"
}

@article{Doneva:2017bvd,
    author = "Doneva, Daniela D. and Yazadjiev, Stoytcho S.",
    title = "{New Gauss-Bonnet Black Holes with Curvature-Induced Scalarization in Extended Scalar-Tensor Theories}",
    eprint = "1711.01187",
    archivePrefix = "arXiv",
    primaryClass = "gr-qc",
    doi = "10.1103/PhysRevLett.120.131103",
    journal = "Phys. Rev. Lett.",
    volume = "120",
    number = "13",
    pages = "131103",
    year = "2018"
}

@article{Silva:2017uqg,
    author = "Silva, Hector O. and Sakstein, Jeremy and Gualtieri, Leonardo and Sotiriou, Thomas P. and Berti, Emanuele",
    title = "{Spontaneous scalarization of black holes and compact stars from a Gauss-Bonnet coupling}",
    eprint = "1711.02080",
    archivePrefix = "arXiv",
    primaryClass = "gr-qc",
    doi = "10.1103/PhysRevLett.120.131104",
    journal = "Phys. Rev. Lett.",
    volume = "120",
    number = "13",
    pages = "131104",
    year = "2018"
}

@article{Antoniou_2018_1,
   title={Black-hole solutions with scalar hair in Einstein-scalar-Gauss-Bonnet theories},
   eprint = "1711.07431",
   archivePrefix = "arXiv",
   primaryClass = "hep-th",
   volume={97},
   ISSN={2470-0029},
   url={http://dx.doi.org/10.1103/PhysRevD.97.084037},
   DOI={10.1103/physrevd.97.084037},
   number={8},
   journal={Physical Review D},
   publisher={American Physical Society (APS)},
   author={Antoniou, G. and Bakopoulos, A. and Kanti, P.},
   year={2018},
   month=apr }

@article{Kokkotas_2017,
   title={Analytical approximation for the Einstein-dilaton-Gauss-Bonnet black hole metric},
   volume={96},
   ISSN={2470-0029},
   url={http://dx.doi.org/10.1103/PhysRevD.96.064004},
   DOI={10.1103/physrevd.96.064004},
   number={6},
   journal={Physical Review D},
   publisher={American Physical Society (APS)},
   author={Kokkotas, K. D. and Konoplya, R. A. and Zhidenko, A.},
   year={2017},
   month=sep }

@article{Blazquez-Salcedo:2018jnn,
    author = "Bl\'azquez-Salcedo, Jose Luis and Doneva, Daniela D. and Kunz, Jutta and Yazadjiev, Stoytcho S.",
    title = "{Radial perturbations of the scalarized Einstein-Gauss-Bonnet black holes}",
    eprint = "1805.05755",
    archivePrefix = "arXiv",
    primaryClass = "gr-qc",
    doi = "10.1103/PhysRevD.98.084011",
    journal = "Phys. Rev. D",
    volume = "98",
    number = "8",
    pages = "084011",
    year = "2018"
}

@article{Minamitsuji:2018xde,
    author = "Minamitsuji, Masato and Ikeda, Taishi",
    title = "{Scalarized black holes in the presence of the coupling to Gauss-Bonnet gravity}",
    eprint = "1812.03551",
    archivePrefix = "arXiv",
    primaryClass = "gr-qc",
    doi = "10.1103/PhysRevD.99.044017",
    journal = "Phys. Rev. D",
    volume = "99",
    number = "4",
    pages = "044017",
    year = "2019"
}

@article{Silva:2018qhn,
    author = "Silva, Hector O. and Macedo, Caio F. B. and Sotiriou, Thomas P. and Gualtieri, Leonardo and Sakstein, Jeremy and Berti, Emanuele",
    title = "{Stability of scalarized black hole solutions in scalar-Gauss-Bonnet gravity}",
    eprint = "1812.05590",
    archivePrefix = "arXiv",
    primaryClass = "gr-qc",
    doi = "10.1103/PhysRevD.99.064011",
    journal = "Phys. Rev. D",
    volume = "99",
    number = "6",
    pages = "064011",
    year = "2019"
}

@article{Blazquez-Salcedo:2020rhf,
    author = "Bl\'azquez-Salcedo, Jose Luis and Doneva, Daniela D. and Kahlen, Sarah and Kunz, Jutta and Nedkova, Petya and Yazadjiev, Stoytcho S.",
    title = "{Axial perturbations of the scalarized Einstein-Gauss-Bonnet black holes}",
    eprint = "2003.02862",
    archivePrefix = "arXiv",
    primaryClass = "gr-qc",
    doi = "10.1103/PhysRevD.101.104006",
    journal = "Phys. Rev. D",
    volume = "101",
    number = "10",
    pages = "104006",
    year = "2020"
}

@article{Blazquez-Salcedo:2020caw,
    author = "Bl\'azquez-Salcedo, Jose Luis and Doneva, Daniela D. and Kahlen, Sarah and Kunz, Jutta and Nedkova, Petya and Yazadjiev, Stoytcho S.",
    title = "{Polar quasinormal modes of the scalarized Einstein-Gauss-Bonnet black holes}",
    eprint = "2006.06006",
    archivePrefix = "arXiv",
    primaryClass = "gr-qc",
    doi = "10.1103/PhysRevD.102.024086",
    journal = "Phys. Rev. D",
    volume = "102",
    number = "2",
    pages = "024086",
    year = "2020"
}

@article{Minamitsuji:2022mlv,
    author = "Minamitsuji, Masato and Takahashi, Kazufumi and Tsujikawa, Shinji",
    title = "{Linear stability of black holes in shift-symmetric Horndeski theories with a time-independent scalar field}",
    eprint = "2201.09687",
    archivePrefix = "arXiv",
    primaryClass = "gr-qc",
    reportNumber = "YITP-22-06, WUCG-22-01",
    doi = "10.1103/PhysRevD.105.104001",
    journal = "Phys. Rev. D",
    volume = "105",
    number = "10",
    pages = "104001",
    year = "2022"
}

@article{East:2020hgw,
    author = "East, William E. and Ripley, Justin L.",
    title = "{Evolution of Einstein-scalar-Gauss-Bonnet gravity using a modified harmonic formulation}",
    eprint = "2011.03547",
    archivePrefix = "arXiv",
    primaryClass = "gr-qc",
    doi = "10.1103/PhysRevD.103.044040",
    journal = "Phys. Rev. D",
    volume = "103",
    number = "4",
    pages = "044040",
    year = "2021"
}

@article{Witek:2018dmd,
    author = "Witek, Helvi and Gualtieri, Leonardo and Pani, Paolo and Sotiriou, Thomas P.",
    title = "{Black holes and binary mergers in scalar Gauss-Bonnet gravity: scalar field dynamics}",
    eprint = "1810.05177",
    archivePrefix = "arXiv",
    primaryClass = "gr-qc",
    doi = "10.1103/PhysRevD.99.064035",
    journal = "Phys. Rev. D",
    volume = "99",
    number = "6",
    pages = "064035",
    year = "2019"
}

@article{Okounkova:2020rqw,
    author = "Okounkova, Maria",
    title = "{Numerical relativity simulation of GW150914 in Einstein dilaton Gauss-Bonnet gravity}",
    eprint = "2001.03571",
    archivePrefix = "arXiv",
    primaryClass = "gr-qc",
    doi = "10.1103/PhysRevD.102.084046",
    journal = "Phys. Rev. D",
    volume = "102",
    number = "8",
    pages = "084046",
    year = "2020"
}

@article{East:2021bqk,
    author = "East, William E. and Ripley, Justin L.",
    title = "{Dynamics of Spontaneous Black Hole Scalarization and Mergers in Einstein-Scalar-Gauss-Bonnet Gravity}",
    eprint = "2105.08571",
    archivePrefix = "arXiv",
    primaryClass = "gr-qc",
    doi = "10.1103/PhysRevLett.127.101102",
    journal = "Phys. Rev. Lett.",
    volume = "127",
    number = "10",
    pages = "101102",
    year = "2021"
}

@article{Corman:2022xqg,
    author = "Corman, Maxence and Ripley, Justin L. and East, William E.",
    title = "{Nonlinear studies of binary black hole mergers in Einstein-scalar-Gauss-Bonnet gravity}",
    eprint = "2210.09235",
    archivePrefix = "arXiv",
    primaryClass = "gr-qc",
    doi = "10.1103/PhysRevD.107.024014",
    journal = "Phys. Rev. D",
    volume = "107",
    number = "2",
    pages = "024014",
    year = "2023"
}

@article{Doneva:2023oww,
    author = "Doneva, Daniela D. and Arest{\'e} Sal{\'o}, Llibert and Clough, Katy and Figueras, Pau and Yazadjiev, Stoytcho S.",
    title = "{Testing the limits of scalar-Gauss-Bonnet gravity through nonlinear evolutions of spin-induced scalarization}",
    eprint = "2307.06474",
    archivePrefix = "arXiv",
    primaryClass = "gr-qc",
    doi = "10.1103/PhysRevD.108.084017",
    journal = "Phys. Rev. D",
    volume = "108",
    number = "8",
    pages = "084017",
    year = "2023"
}

@misc{AresteSalo:2025sxc,
    author = "Arest{\'e} Sal{\'o}, Llibert and Doneva, Daniela D. and Clough, Katy and Figueras, Pau and Yazadjiev, Stoytcho S.",
    title = "{Challenges in the nonlinear evolution of unequal mass binaries in sGB gravity}",
    eprint = "2507.13046",
    archivePrefix = "arXiv",
    primaryClass = "gr-qc",
    month = "7",
    year = "2025"
}

@article{Carson:2019fxr,
    author = "Carson, Zack and Seymour, Brian C. and Yagi, Kent",
    title = "{Future prospects for probing scalar{\textendash}tensor theories with gravitational waves from mixed binaries}",
    eprint = "1907.03897",
    archivePrefix = "arXiv",
    primaryClass = "gr-qc",
    doi = "10.1088/1361-6382/ab6a1f",
    journal = "Class. Quant. Grav.",
    volume = "37",
    number = "6",
    pages = "065008",
    year = "2020"
}

@article{Thornburg:2006zb,
    author = "Thornburg, Jonathan",
    title = "{Event and apparent horizon finders for 3+1 numerical relativity}",
    eprint = "gr-qc/0512169",
    archivePrefix = "arXiv",
    reportNumber = "AEI-2005-184",
    doi = "10.12942/lrr-2007-3",
    journal = "Living Rev. Rel.",
    volume = "10",
    pages = "3",
    year = "2007"
}

@article{Masso:1998fi,
    author = "Masso, Joan and Seidel, Edward and Suen, Wai-Mo and Walker, Paul",
    title = "{Event horizons in numerical relativity 2.: Analyzing the horizon}",
    eprint = "gr-qc/9804059",
    archivePrefix = "arXiv",
    doi = "10.1103/PhysRevD.59.064015",
    journal = "Phys. Rev. D",
    volume = "59",
    pages = "064015",
    year = "1999"
}

@article{Libson:1994dk,
    author = "Libson, Joseph and Masso, Joan and Seidel, Edward and Suen, Wai-Mo and Walker, Paul",
    title = "{Event horizons in numerical relativity. 1: Methods and tests}",
    eprint = "gr-qc/9412068",
    archivePrefix = "arXiv",
    doi = "10.1103/PhysRevD.53.4335",
    journal = "Phys. Rev. D",
    volume = "53",
    pages = "4335--4350",
    year = "1996"
}

@article{Maselli:2020zgv,
    author = "Maselli, Andrea and Franchini, Nicola and Gualtieri, Leonardo and Sotiriou, Thomas P.",
    title = "{Detecting scalar fields with Extreme Mass Ratio Inspirals}",
    eprint = "2004.11895",
    archivePrefix = "arXiv",
    primaryClass = "gr-qc",
    doi = "10.1103/PhysRevLett.125.141101",
    journal = "Phys. Rev. Lett.",
    volume = "125",
    number = "14",
    pages = "141101",
    year = "2020"
}

@article{Davis:1971gg,
    author = "Davis, M. and Ruffini, R. and Press, W. H. and Price, R. H.",
    title = "{Gravitational radiation from a particle falling radially into a schwarzschild black hole}",
    doi = "10.1103/PhysRevLett.27.1466",
    journal = "Phys. Rev. Lett.",
    volume = "27",
    pages = "1466--1469",
    year = "1971"
}

@misc{Capuano:2026lhs,
    author = "Capuano, Lodovico and Arest{\'e} Sal{\'o}, Llibert and Doneva, Daniela D. and Yazadjiev, Stoytcho S. and Barausse, Enrico",
    title = "{Dynamical hair growth in black hole binaries in Einstein-scalar-Gauss-Bonnet gravity}",
    eprint = "2602.02650",
    archivePrefix = "arXiv",
    primaryClass = "gr-qc",
    month = "2",
    year = "2026"
}

@article{Booth:2020qhb,
    author = "Booth, Ivan and Hennigar, Robie A. and Mondal, Saikat",
    title = "{Marginally outer trapped surfaces in the Schwarzschild spacetime: Multiple self-intersections and extreme mass ratio mergers}",
    eprint = "2005.05350",
    archivePrefix = "arXiv",
    primaryClass = "gr-qc",
    doi = "10.1103/PhysRevD.102.044031",
    journal = "Phys. Rev. D",
    volume = "102",
    number = "4",
    pages = "044031",
    year = "2020"
}

@article{Pook-Kolb:2021gsh,
    author = "Pook-Kolb, Daniel and Hennigar, Robie A. and Booth, Ivan",
    title = "{What Happens to Apparent Horizons in a Binary Black Hole Merger?}",
    eprint = "2104.10265",
    archivePrefix = "arXiv",
    primaryClass = "gr-qc",
    doi = "10.1103/PhysRevLett.127.181101",
    journal = "Phys. Rev. Lett.",
    volume = "127",
    number = "18",
    pages = "181101",
    year = "2021"
}

@article{Ferrari:1984zz,
    author = "Ferrari, Valeria and Mashhoon, Bahram",
    title = "{New approach to the quasinormal modes of a black hole}",
    doi = "10.1103/PhysRevD.30.295",
    journal = "Phys. Rev. D",
    volume = "30",
    pages = "295--304",
    year = "1984"
}

@article{Mashhoon:1985cya,
    author = "Mashhoon, Bahram",
    title = "{Stability of charged rotating black holes in the eikonal approximation}",
    doi = "10.1103/PhysRevD.31.290",
    journal = "Phys. Rev. D",
    volume = "31",
    number = "2",
    pages = "290--293",
    year = "1985"
}

@article{Cardoso:2008bp,
    author = "Cardoso, Vitor and Miranda, Alex S. and Berti, Emanuele and Witek, Helvi and Zanchin, Vilson T.",
    title = "{Geodesic stability, Lyapunov exponents and quasinormal modes}",
    eprint = "0812.1806",
    archivePrefix = "arXiv",
    primaryClass = "hep-th",
    doi = "10.1103/PhysRevD.79.064016",
    journal = "Phys. Rev. D",
    volume = "79",
    number = "6",
    pages = "064016",
    year = "2009"
}

@article{Yunes:2011we,
    author = "Yunes, Nicolas and Stein, Leo C.",
    title = "{Non-Spinning Black Holes in Alternative Theories of Gravity}",
    eprint = "1101.2921",
    archivePrefix = "arXiv",
    primaryClass = "gr-qc",
    doi = "10.1103/PhysRevD.83.104002",
    journal = "Phys. Rev. D",
    volume = "83",
    pages = "104002",
    year = "2011"
}

@article{Gross:1986mw,
    author = "Gross, David J. and Sloan, John H.",
    title = "{The Quartic Effective Action for the Heterotic String}",
    reportNumber = "NSF-ITP-87-02",
    doi = "10.1016/0550-3213(87)90465-2",
    journal = "Nucl. Phys. B",
    volume = "291",
    pages = "41--89",
    year = "1987"
}

@article{Metsaev:1987zx,
    author = "Metsaev, R. R. and Tseytlin, Arkady A.",
    title = "{Order alpha-prime (Two Loop) Equivalence of the String Equations of Motion and the Sigma Model Weyl Invariance Conditions: Dependence on the Dilaton and the Antisymmetric Tensor}",
    reportNumber = "PRINT-87-0184 (LEBEDEV)",
    doi = "10.1016/0550-3213(87)90077-0",
    journal = "Nucl. Phys. B",
    volume = "293",
    pages = "385--419",
    year = "1987"
}

@article{Antoniou:2017acq,
    author = "Antoniou, G. and Bakopoulos, A. and Kanti, P.",
    title = "{Evasion of No-Hair Theorems and Novel Black-Hole Solutions in Gauss-Bonnet Theories}",
    eprint = "1711.03390",
    archivePrefix = "arXiv",
    primaryClass = "hep-th",
    doi = "10.1103/PhysRevLett.120.131102",
    journal = "Phys. Rev. Lett.",
    volume = "120",
    number = "13",
    pages = "131102",
    year = "2018"
}

@article{Nojiri:2005vv,
    author = "Nojiri, Shin'ichi and Odintsov, Sergei D. and Sasaki, Misao",
    title = "{Gauss-Bonnet dark energy}",
    eprint = "hep-th/0504052",
    archivePrefix = "arXiv",
    reportNumber = "YITP-05-14",
    doi = "10.1103/PhysRevD.71.123509",
    journal = "Phys. Rev. D",
    volume = "71",
    pages = "123509",
    year = "2005"
}

@article{Weinberg:2008hq,
    author = "Weinberg, Steven",
    title = "{Effective Field Theory for Inflation}",
    eprint = "0804.4291",
    archivePrefix = "arXiv",
    primaryClass = "hep-th",
    reportNumber = "UTTG-01-08",
    doi = "10.1103/PhysRevD.77.123541",
    journal = "Phys. Rev. D",
    volume = "77",
    pages = "123541",
    year = "2008"
}

@article{Burgess:2003jk,
    author = "Burgess, C. P.",
    title = "{Quantum gravity in everyday life: General relativity as an effective field theory}",
    eprint = "gr-qc/0311082",
    archivePrefix = "arXiv",
    doi = "10.12942/lrr-2004-5",
    journal = "Living Rev. Rel.",
    volume = "7",
    pages = "5--56",
    year = "2004"
}

@article{Wong:2022wni,
    author = "Wong, Leong Khim and Herdeiro, Carlos A. R. and Radu, Eugen",
    title = "{Constraining spontaneous black hole scalarization in scalar-tensor-Gauss-Bonnet theories with current gravitational-wave data}",
    eprint = "2204.09038",
    archivePrefix = "arXiv",
    primaryClass = "gr-qc",
    doi = "10.1103/PhysRevD.106.024008",
    journal = "Phys. Rev. D",
    volume = "106",
    number = "2",
    pages = "024008",
    year = "2022"
}

@article{Yordanov:2024lfk,
    author = "Yordanov, Petar Y. and Staykov, Kalin V. and Yazadjiev, Stoytcho S. and Doneva, Daniela D.",
    title = "{The power of binary pulsars in testing Gauss-Bonnet gravity}",
    eprint = "2402.06305",
    archivePrefix = "arXiv",
    primaryClass = "gr-qc",
    doi = "10.1051/0004-6361/202449679",
    journal = "Astron. Astrophys.",
    volume = "687",
    pages = "A17",
    year = "2024"
}

@article{Jaramillo:2023day,
    author = "Jaramillo, Jos{\'e} Luis and Krishnan, Badri and Sopuerta, Carlos F.",
    title = "{Universality in binary black hole dynamics: An integrability conjecture}",
    eprint = "2305.08554",
    archivePrefix = "arXiv",
    primaryClass = "gr-qc",
    doi = "10.1142/S0218271823420051",
    journal = "Int. J. Mod. Phys. D",
    volume = "32",
    number = "14",
    pages = "2342005",
    year = "2023"
}

@article{Pook-Kolb:2019ssg,
    author = "Pook-Kolb, Daniel and Birnholtz, Ofek and Krishnan, Badri and Schnetter, Erik",
    title = "{Self-intersecting marginally outer trapped surfaces}",
    eprint = "1907.00683",
    archivePrefix = "arXiv",
    primaryClass = "gr-qc",
    doi = "10.1103/PhysRevD.100.084044",
    journal = "Phys. Rev. D",
    volume = "100",
    number = "8",
    pages = "084044",
    year = "2019"
}

@misc{Pook-Kolb:2020zhm,
    author = "Pook-Kolb, Daniel and Birnholtz, Ofek and Jaramillo, Jos{\'e} Luis and Krishnan, Badri and Schnetter, Erik",
    title = "{Horizons in a binary black hole merger I: Geometry and area increase}",
    eprint = "2006.03939",
    archivePrefix = "arXiv",
    primaryClass = "gr-qc",
    month = "6",
    year = "2020"
}

@misc{Pook-Kolb:2020jlr,
    author = "Pook-Kolb, Daniel and Birnholtz, Ofek and Jaramillo, Jos{\'e} Luis and Krishnan, Badri and Schnetter, Erik",
    title = "{Horizons in a binary black hole merger II: Fluxes, multipole moments and stability}",
    eprint = "2006.03940",
    archivePrefix = "arXiv",
    primaryClass = "gr-qc",
    month = "6",
    year = "2020"
}

\end{document}